%% file: main.tex
\documentclass[useAMS,usenatbib,fleqn]{mnras}
\usepackage{psfrag}
\usepackage{graphicx}	
\usepackage{amsmath}	
\usepackage{amssymb}	
\usepackage{multicol}        
\usepackage{bm}		
\usepackage{hyperref}
\usepackage{xcolor}

\usepackage[T1]{fontenc}
\usepackage{ae,aecompl}

\usepackage{breqn}

\def\kk{\mathbf{k}}
\def\ka{\mathbf{k_1}}
\def\kb{\mathbf{k_2}}
\def\kc{\mathbf{k_3}}

\def\T{\mathcal{T}}

\def\zh{\mathbf{\hat{z}}}

\def\ph{\mathbf{\hat{p}}}

\def\RB{\bar{R}}
\def\nn{\nonumber}

\def\S#1{S_{#1}}
\def\T#1{T_{#1}}
\def\SB#1#2#3{[\bar{S}_{#3}]^{#1}_{#2}}
\def\TB#1#2#3{[\bar{T}_{#3}]^{#1}_{#2}}
\def\bl{b_1}
\def\gn{\gamma_2}
\def\bto{\beta_1}
\defcitealias{Bharadwaj:2020wkc}{Paper~I}

\title[ Redshift Space  Bispectrum II: Induced Non-Gaussianity]{Quantifying the Redshift Space Distortion of the Bispectrum II: Induced Non-Gaussianity at Second Order Perturbation}
\author[A. Mazumdar, S. Bharadwaj, D. Sarkar]{ Arindam Mazumdar$^{1}$\thanks{\href{mailto:arindam.mazumdar@iitkgp.ac.in}{arindam.mazumdar@iitkgp.ac.in}}, Somnath Bharadwaj$^{1,2}$\thanks{\href{mailto:somnath@phy.iitkgp.ernet.in}{somnath@phy.iitkgp.ernet.in}},   Debanjan Sarkar$^{1,3}$\thanks{\href{mailto:debanjan@post.bgu.ac.il}{debanjan@post.bgu.ac.il}}
\\
$^{1}$Centre for Theoretical Studies, Indian Institute of Technology Kharagpur, Kharagpur - 721302, India\\
$^{2}$Department of Physics, Indian Institute of Technology Kharagpur, Kharagpur - 721302, India\\
$^{3}$Department of Physics, Ben-Gurion University of the Negev, Be’er Sheva - 84105, Israel}
\date{}

\pubyear{2020}

\begin{document}
\label{firstpage}
\pagerange{\pageref{firstpage}--\pageref{lastpage}}
\maketitle

\begin{abstract}
The anisotrpy of the redshift space bispectrum $B^s(\ka,\kb,\kc)$, which 
contains a wealth of cosmological information, is completely quantified using 
multipole moments $\bar{B}^m_{\ell}(k_1,\mu,t)$ where $k_1$, the length of the 
largest side, and $(\mu,t)$ respectively quantify the size and shape of the 
triangle $(\ka,\kb,\kc)$. We present  analytical expressions for all the 
multipoles which are predicted to be non-zero ($\ell \le 8, m \le 6$ ) at 
second 
order perturbation theory. The multipoles also depend on $\bto,\bl$ and $\gn,$ 
which quantify the linear redshift distortion parameter, linear bias and 
quadratic bias respectively. Considering triangles of all possible shapes, we 
analyse the shape dependence of all of the
multipoles holding $k_1=0.2 \, {\rm Mpc}^{-1}, \bto=1, \bl=1$ and $\gn=0$ fixed.
The monopole $\bar{B}^0_0$, which is positive everywhere, is minimum for 
equilateral 
triangles. $\bar{B}_0^0$ increases towards  linear triangles, and is maximum  
for 
linear triangles close to the squeezed limit.  Both $\bar{B}^0_{2}$ and 
$\bar{B}^0_4$ are 
similar to $\bar{B}^0_0$, however the quadrupole $\bar{B}^0_2$ exceeds 
$\bar{B}^0_0$ over a 
significant range of shapes. The other multipoles, many of which become 
negative, have magnitudes smaller than $\bar{B}^0_0$. In most cases the maxima 
or 
minima, or both, occur very close to the squeezed limit. 
$\mid \bar{B}^m_{\ell} \mid $ is found to decrease rapidly if $\ell$ or $m$ are 
increased. The shape 
dependence shown here is characteristic of non-linear gravitational clustering. 
Non-linear bias, if present, will lead to a different shape dependence. 
\end{abstract}




\section{Introduction}
\label{sec:intro}

The simplest models of inflation  predict the primordial density fluctuations which seed the large-scale structures in the Universe to be a Gaussian random field \citep{baumann09-inflation}, 
 Cosmic Microwave Background (CMB) measurements 
 \citep{fergusson12-PNG,oppizzi18-PNG, planck18png, shiraishi19-PNG} and galaxy surveys \citep{feldman01-bispec, scoccimarro04-PNG, ligouri10-PNG, ballardini19png} have been used to place tight constraints on primordial non-Gaussianity. These  density fluctuations are however predicted to become non-Gaussian as they evolve (induced non-Gaussianity; \citealt{fry84-bispec}) 
	due to the non-linear growth and non-linear biasing.  It  is therefore  
	necessary to consider higher order statistics,  the three-point correlation function or its Fourier conjugate the bispectrum being  the lowest order statistic sensitive to non-Gaussianity. 
Second order perturbation theory predicts \citep{matarrese97} that measurements of the bispectrum in the weakly non-linear regime can be used to determine the bias parameters, and this has been employed in the galaxy surveys to quantify the galaxy bias parameters \citep{feldman01-bispec, scoccimarro01-IRAS-bispec, verde02-2DF, nishimichi07, gilmartin15-bispec}. Further, the measurements of bispectrum enable us to lift the degeneracy between $\Omega_m$ (which appears in $f(\Omega_m)$) 
	and $b_1$, something  which is not possible by considering only the power spectrum \citep{scoccimarro99-RS-bispec}.

Redshift space distortion(RSD) is a prominent feature in redshift surveys which probe the  large scale structures observed in  the Universe. At small length-scales  random motions make the structures appear elongated along line of sight (LoS) leading to the well known Finger of God (FoG) effect \citep{jackson72-FoG}. At large length-scales RSD causes the  over-dense regions to appear more over-dense and the  under-dense  regions to appear  more under-dense along the LoS \citep{kaiser87}. At large length-scales where linear perturbation theory may be assumed to hold, 
the redshift space power spectrum of a linearly biased tracer, $P^s(\kk_1)$, is related to the real space power spectrum $P^r(k_1)$ as $P^s(\kk_1)=(1+\bto \mu_1^2)^2 P^r(k_1)$  \citep{kaiser87} where $\bto=f/b_1$ is the linear redshift distortion parameter. Here $f$ is logarithmic derivative of the growth rate	of density perturbation in linear theory and this is a function of cosmological matter density parameter $\Omega_m$. $b_1$ is linear bias factor and $\mu_1=\zh \cdot \ka /k_1$ where $\zh$ points towards the LoS direction.
The redshift space power spectrum $P^s(\kk_1)$ is anisotropic {\it i.e.} it depends on how $\ka$ is oriented with respect to $\zh$. This anisotropy can be quantified by expanding $P^s(\kk_1)$  in terms of Legendre polynomials in $\mu_1$, the reader is referred to an extensive review \citep{hamilton-98-rsd-review} of linear RSD for details. The anisotropy of the redshift space 
power-spectrum contains a rich wealth of cosmological information. 
For example, the parameter $f$ can be estimated \citep{loveday96, peacock01, hawkins03, guzzo-pierleoni08}, the total mass of massive neutrinos can be constrained \citep{hu98-neutrino,Upadhye:2017hdl}, and  dark energy theories and modified gravity theories can be tested \citep{linder08-RSD-GR, song09-RSD, de_la_torre16-gravity-test-from-RSD-and-lensing, johnson-blake-16-modified-gravity-using-galaxy-peculiar-velocities, mueller18-BOSS-RSD}.

 In this paper we consider the redshift space bispectrum $B^s(\ka,\kb,\kc)$ which is induced at second order perturbation theory starting from Gaussian initial conditions.  In addition to the size and shape of the triangle $(\ka,\kb,\kc)$, the redshift space bispectrum also depends on how the three vectors are oriented with respect to $\zh$.  	Like the power spectrum, the anisotropy of  $B^s(\ka,\kb,\kc)$ 
 contains a wealth of cosmological information, and  it is important to accurately model and quantify this. \citet{Hivon1995} and  \citet{Verde1998} have calculated the bispectrum in redshift space. However,  the focus in these works has been on  measuring the large scale bias and the cosmological parameters, and they have not quantified the RSD anisotropy.
\citet{scoccimarro99-RS-bispec} have quantified the  anisotropy of the redshift space  bispectrum by decomposing it into spherical harmonics.  However, they have only considered the monopole and one  quadrupole component $(\ell=2,m=0)$. 
\citet{Hashimoto2017} also have considered a  single quadrupole component of the redshift space bispectrum. The works mentioned above are all   use non-linear perturbation theory, and  their results are extensively validated using large cosmological N-body simulations. However, these works do not carry out a complete analysis of the anisotropy,  and are 
 restricted  to a single quadrupole component and a very limited  set of triangle configurations.
\citet{Nan} have presented approximate analytical expressions based on the halo model for the  higher angular multipole moments up to $\ell=4$, and they have analysed these for a limited  set of triangle configurations. 
\citet{Yankelevich} and \citet{Gualdi} forecast cosmological parameter constraints  including the redshift space bispectrum and power spectrum. \citet{desjacques18-RSD-EFT} have used effective field theory  to accurately model 
 the redshift space bispectrum on mildly non-linear scales.  \citealt{Clarkson:2018dwn} and \cite{deWeerd:2019cae} have recently shown that relativistic effects will introduce a dipole anisotropy in the redshift space bispectrum on  very large length-scales. 
\citet{Slepian:2016weg}, \citet{Slepian2018} present a technique to quantify the redshift space three-point correlation function by expanding it in terms of products of two spherical harmonics, whereas  
\citet{Sugiyama2019a} have proposed  a tri-polar spherical harmonic decomposition to quantify the anisotropy of the redshift space bispectrum   which they  demonstrate by applying it to the Baryon Oscillation Spectroscopic Survey (BOSS) Data Release 12. The last two works  mentioned here present very efficient computational techniques
for quantifying the three point statistics in  large galaxy surveys. 
	
In a recent work \citep{Bharadwaj:2020wkc} (hereafter referred to   as \citetalias{Bharadwaj:2020wkc})  we have proposed  a new technique to quantify  the anisotropy of the redshift space bispectrum.  We have decomposed the redshift space bispectrum in spherical harmonics which completely quantify the anisotropy. We illustrate this by  considering the  linear RSD  of the bispectrum arising from primordial non-Gaussianity. Only the first four even  $\ell$  multipoles (up to $\ell=6,m=4$)  are found to 
be  non-zero, and we have presented  explicit analytical expressions for  these.
 We find that the ratio of the different  multipole moments to the real space bispectrum are cubic polynomials in  
 $\beta_1$ the linear redshift distortion parameter. The  coefficients of these polynomials depend only on the 
 shape of the triangle. We have analysed all the non-zero multipole moments for triangles of all possible shapes.  If measured in future, the various  multipole moments  of the bispectrum of primordial non-Gaussianity hold the potential of constraining $\beta_1$.  The results presented in \citetalias{Bharadwaj:2020wkc}  are also important to  constrain  $f_{\text{NL}}$ using redshift surveys. 

In the present paper we have applied the formalism developed in \citetalias{Bharadwaj:2020wkc} to quantify the anisotropy of the induced redshift space at second order perturbation theory. We present explicit analytical formulas  for all the non-zero multipoles moments of the induced redshift space bispectrum .  Considering triangles of all possible shapes, we analyse the variation of different multipole moments with the shape of the triangle. A brief outline of the rest of the paper follows.  In Section~\ref{sec:form} we briefly summarize some of the salient features of the formalism developed in \citetalias{Bharadwaj:2020wkc}. 
We first apply the formalism (Section~\ref{sec:bispec-real})  to quantify and analyse the shape dependence of the real space bispectrum.   In Section~\ref{sec:bispec} we develop  the methodology and notation to  calculate  the   multipole moments of the induced redshift space bispectrum. The final expression for the $\ell,m$ multipole refers to  three terms namely $R$ whose multipole moments are presented in \citetalias{Bharadwaj:2020wkc}, $S$ and $T$ whose multipole moments are presented in Appendix~\ref{sec:a1} and \ref{sec:a2}   respectively.  We present the results in section~\ref{sec:results}, whereas 
section~\ref{sec:summ} presents Summary and Discussion. 

We have used the $\Lambda$CDM power spectrum generated by the 
Blotzmann code CLASS (\citealt{Lesgourgues:2011re,Blas:2011rf}) with the cosmological parameters fixed from Planck 2015 
results~\citealt{planck-collaboration15}.

\section{Formalism}
\label{sec:form}
\begin{figure}
    \centering
    \includegraphics[width=5cm]{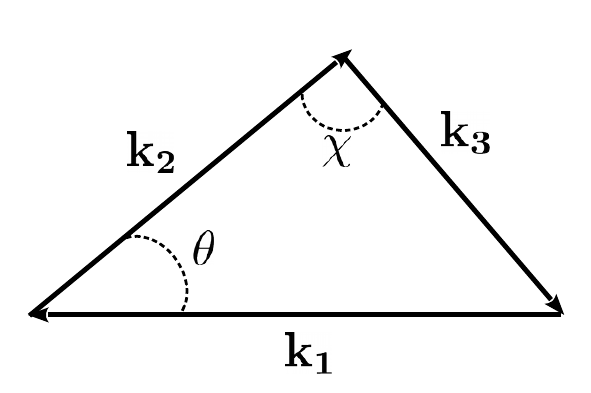}
    \caption{This shows a triangle $(\ka,\kb,\kc)$ which is used to define various parameters used here. }
    \label{fig:triangle}
\end{figure}

The   bispectrum is defined  as
\begin{equation}
	B(\ka,\kb,\kc)=V^{-1}\,  \langle \Delta(\ka) \Delta(\kb) \Delta(\kc) \rangle \,,
	\label{eq:a1}
\end{equation}
where the $\Delta(\kk)$s  refer to  the Fourier components    of  the    density contrast.    The three vectors involved in the bispectrum are constrained to form a closed triangle, $i.e.\, \ka+\kb+\kc = 0$ (Figure \ref{fig:triangle}). We label  the three sides of the triangle such that  $k_1 \ge k_2 \ge k_3$ where $k_1=\mid \ka \mid$,  etc. The real space bispectrum  $B^r(\ka,\kb,\kc)$ 
is independent of how the triangle is oriented in space, and it depends only on the size and shape of the triangle.
 Following \citetalias{Bharadwaj:2020wkc}, we use the length of the largest side $k_1$  to parameterize the size of the triangle, and we parameterize the shape using $\mu=\cos \theta =-\ka\cdot \kb/(k_1 k_2)$  which is the   cosine of the angle between $- \kb$ and $\ka$,  and $t=k_2/k_1$  which is the ratio of the second largest side to the largest side. The values of  $\mu$ and $t$ are restricted to the range 
\begin{equation}
0.5 \le t,\mu \le 1 \, \, {\rm and}\, \,2  \mu t \ge 1   \, ,  
\end{equation}
and these uniquely specify the shapes of all possible triangles. Figure~2 of \citetalias{Bharadwaj:2020wkc} provides a detailed description of the  triangle shape corresponding to different values of the parameters $(\mu,t)$. For completeness, we summarize this later in this paper when we analyze  the shape dependence of the real space bispectrum. 
In the subsequent discussion we use $B^r(k_1,\mu,t)$ to parameterize the shape and size dependence of the real space bispectrum.  Further, it is convenient  to use the notation $s=k_3/k_1=\sqrt{t^2-2 \mu t +1}$. 

The redshift space bispectrum $B^s(\ka,\kb,\kc)$, unlike its real space counterpart, depends 
on the orientations of the triangles with respect to the line of sight (LoS)  direction $\zh$. This anisotropy or orientation dependence arises through  $\mu_1,\mu_2,\mu_3$  which are respectively the cosine of the angles between $\ka,\kb,\kc$  and  $\zh$.
It is necessary to consider a triangle with  a  fixed size and shape, and vary its orientation in order to quantify the anisotropy arising from RSD. As discussed in \citetalias{Bharadwaj:2020wkc}, this can be achieved by applying rigid body rotations to a triangle whose size and shape are fixed. It is then possible to parameterize  the  redshift space bispectrum as $B^s(\alpha,\beta,\gamma,k_1,\mu,t)$ where $(\alpha,\beta,\gamma)$ are three angles needed to parameterize the rigid body rotations. However, for the analytical estimates presented in \citetalias{Bharadwaj:2020wkc} and continued in this paper, it suffices to parameterize  the  redshift space bispectrum as $B^s(\ph,k_1,\mu,t)$ 
where $\ph$ is a unit vector which can vary over  all possible directions. We have 
\begin{eqnarray}
\mu_1 &=& p_z \nn\\
\mu_2 &=&-\mu p_z + \sqrt{1-\mu^2} p_x \nn\\
\mu_3 &=& -s^{-1} [(1-t \mu) p_z + t \sqrt{1-\mu^2} p_x]
\end{eqnarray}
where $s$ is a function of $(\mu,t)$ introduced earlier. 

We quantify the anisotropy of  the redshift space bispectrum $B^s(\ph,k_1,\mu,t)$  using the different multipole moments defined as 
\begin{equation}\label{eq:b9}
\bar{B}^{m}_{\ell}(k_1,\mu,t) =\sqrt{\frac{(2 \ell +1)}{4 \pi}} \int [Y^m_{\ell}(\ph)]^* B^s(\ph,k_1,\mu,t) \, d\Omega_{\ph} 
\end{equation}
where $Y^m_{\ell}(\ph)$ are the  spherical harmonics and the $ d\Omega_{\ph}$
integral is over all possible directions of $\ph$ which subtends $4 \, \pi$ steradians. The reader is referred to \citetalias{Bharadwaj:2020wkc} regarding the choice of normalization and other details for eq.~(\ref{eq:b9}).

The odd multipoles $(\ell=1,3,5,...)$  are all zero  in the plane parallel approximation. We further have  
$ \bar{B}^{-m}_{\ell}(k_1,\mu,t) = (-1)^m \bar{B}^{m}_{\ell}(k_1,\mu,t)$, and  $\bar{B}^{m}_{\ell}(k_1,\mu,t)$ are all real. Considering  linear triangles $(\mu=1)$,   the anisotropy is completely quantified by the $m=0$ multipole moment and  the multipole moments with $m \neq 0$ are all zero. We however need both the $m=0$ and $m \neq 0$ multipole moments to completely quantify the anisotropy of $B^s(\ph,k_1,\mu,t)$ when the three vectors $\ka,\kb$ and $\kc$ are not aligned  {\it i.e.} $(\mu <1)$.

\section{Induced real space bispectrum}\label{sec:bispec-real}
Considering Gaussian initial conditions, non-Gaussianity arises due to the non-linear evolution of the density perturbations \citep{fry84-bispec}. The bispectrum  of induced  non-Gaussianity at second order perturbation theory \citep{scoccimarro98-NL-bispec} is given by 
\begin{dmath}
B^r(\ka,\kb,\kc) = 2 \bl^{-1}\left[F_2(\ka,\kb)+{\gn \over 2}\right]P^r(k_1) P^r(k_2) + {\rm cyc ...}
\label{eq:b1}
\end{dmath}
where $\bl$ is the linear bias, $\gn=b_2/b_1$ parameterizes the quadratic bias $b_2$ in terms of $b_1$, $P^r(k)$ is the real space  power spectrum of the tracer for which the bispectrum is being calculated. Here  $P^r(k)=\bl^2 \, P_m(k)$ where $P_m(k)$ is the matter power spectrum,  and 
\begin{equation}
    F_2(\ka,\kb)={5\over 7}+{\ka\cdot \kb\over 2}\left({1\over k_1^2}+{1\over k_2^2}\right)+{2\over7}{(\ka \cdot \kb)^2\over k_1^2 k_2^2}
\label{eq:FF2}
\end{equation}
are the kernels appearing in the second order perturbation for the density contrast (\citealt{Goroff:1986ep}).  
 The various terms $F_2(\ka,\kb), F_2(\kb,\kc), F_2(\kc,\ka)$ are dimensionless functions which depend on the shape of the triangle $(\ka,\kb,\kc)$, and these have very simple algebraic expressions in terms of $(\mu,t)$. Using the compact notation $F_{12}(\mu,t)=F_2(\ka,\kb)$, etc.  we can express these as 
\begin{eqnarray}
F_{12}(\mu,t) &=& \frac{1}{14} \left(4 \mu ^2-7 \mu  t-\frac{7 \mu }{t}+10\right)\, ,\nn\\
F_{23}(\mu,t)&=&\frac{7 \mu +\left(3-10 \mu ^2\right) t}{14 t s^2}\, ,\nn\\
F_{31}(\mu,t)&=&\frac{t^2 \left(-10 \mu ^2+7 \mu  t+3\right)}{14 s^2} \,.
\label{eq:b3}
\end{eqnarray}

Considering $b_1$ and $\gn$ as independent parameters in eq.~(\ref{eq:b1}),  we see that changing $b_1$  scales the bispectrum  irrespective   of the shape and size of the triangle. In contrast, $\gn$ occurs in combination with  $F_{12}(\mu,t)$, etc. whose values depend on the shape of the triangle.  
It has been proposed \citep{verde98-RS-bispec} that the shape dependence of the induced bispectrum (eq.~\ref{eq:b1}) can be used to independently determine $\bl$ and $\gn$ from the measured bispectrum. 

 In order to  analyze  the theoretical predictions presented here,  following \citet{fry84-bispec} we define a dimensionless bispectrum 
\begin{equation}
    Q^r(k_1,\mu,t)= {\bl \, B^r(k_1,\mu,t)\over 3 [P^{r}(k_1)]^2}\, . 
\label{eq:b4}
\end{equation} 
In addition to being dimensionless,  this  has the added advantage of eliminating the $b_1$ dependence in eq.~(\ref{eq:b1})  which just introduces an overall scaling of the bispectrum. 
Note   that  our $Q^r(k_1,\mu,t)$ is somewhat different  from the dimensionless three point hierarchical amplitude $Q$ used in several earlier works including  \citet{fry84-bispec}. 

\begin{figure}
 \includegraphics[width=0.5\textwidth]{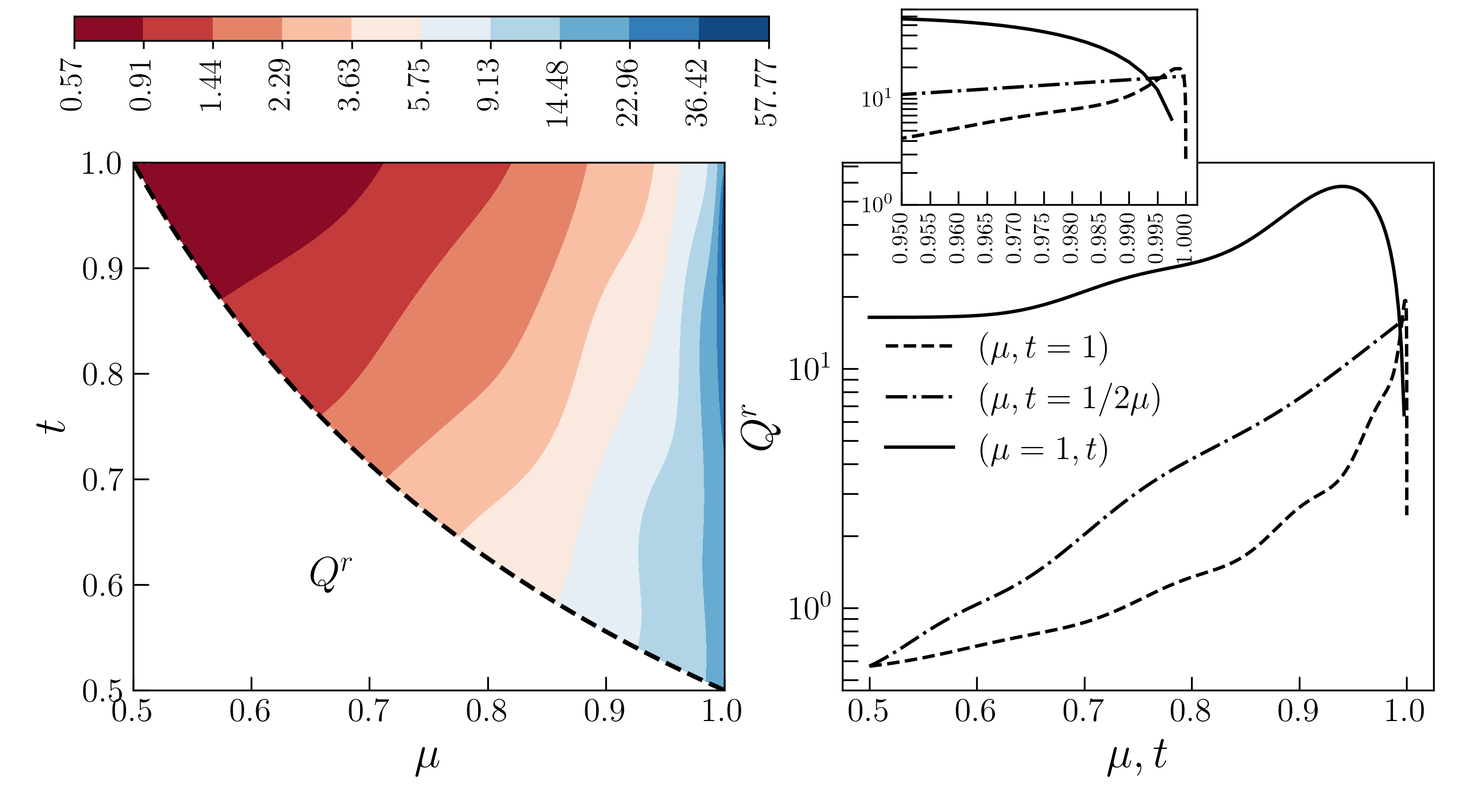}
    \caption{Left panel: $Q^r(\mu,t)$ at $k_1=0.2 \, {\rm Mpc}^{-1}$for the reference model where  $\bl=1$ and $\gn=0$. 
    Right panel:    The different curves show the values of $Q^r(\mu,t)$ along the three boundaries of the allowed region of $(\mu,t)$ space. 
    The $x$-axis corresponds to $\mu$ or $t$ depending on the curve which is being referred to. 
    The bottom  curve shows $Q^r(\mu,1)$ as a  function of $\mu$, this corresponds to L-isosceles triangles (top boundary of left panel). The middle  curve shows $Q^r(\mu,1/(2 \mu))$ as a  function of $\mu$, this corresponds to S-isosceles triangles (bottom boundary of left panel).   The top curve shows $Q^r(1,t)$ as a  function of $t$, this corresponds to linear  triangles (right boundary of left panel). { The inset zooms in on the squeezed limit.}}
    \label{fig:Q_r}
\end{figure}

The left panel of Figure \ref{fig:Q_r} shows $ Q^r(\mu,t)$  which refers to $ Q^r(k_1,\mu,t)$ at $k_1=0.2 \, {\rm Mpc}^{-1}$ for the reference model  where  $b_1=1$ and $\gn=0$, and in the subsequent discussion we shall largely  focus on the $(\mu,t)$  dependence for this fixed value of $k_1$. The value of $k_1$ considered here refers to sufficiently large scales where we may expect second order perturbation theory to provide a reasonably valid descriptions. 
The triangle shapes corresponding to different values of $(\mu,t)$ has been discussed in  \citetalias{Bharadwaj:2020wkc},    we summarize this here.  The right boundary $\mu=1$ corresponds to linear triangles where $\ka$, $-\kb$ and $-\kc$ are parallel (Figure \ref{fig:triangle}). The top right corner $(\mu \rightarrow 1, t  \rightarrow 1)$ and 
and the bottom right corner  $(\mu \rightarrow 1, t  \rightarrow 1/2)$  correspond to squeezed $(\ka=-\kb, \kc \rightarrow 0)$
and stretched $(\kb=\kc=- \ka/2)$ triangles respectively. The top boundary $t=1$ corresponds to L-isosceles triangles where the two larger sides ($\ka$ and $\kb$)  are of equal length, whereas the bottom boundary $2 \mu t =1$   corresponds to S-isosceles triangles where the two smaller  sides ($\kb$ and $\kc$)  are of equal length. The top left corner  $(\mu \rightarrow 1/2, t \rightarrow 1)$ corresponds to equilateral triangles. The diagonal line $\mu=t$ corresponds to right-angle triangles ( $\chi=90^{\circ}$ in Figure \ref{fig:triangle})  while the upper $(t > \mu)$ and lower halves  correspond to acute and obtuse triangles respectively. 
 
Considering $Q^r(\mu,t)$ in the left panel  of Figure  \ref{fig:Q_r}  we see that the value is minimum at 
the top left corner $(\mu,t)=(1/2,1)$ which corresponds to  equilateral triangles where 
 \begin{equation}
 Q^r(1/2,1)=\left({4\over7} + \gn\right) 
 \label{eq:b5}
 \end{equation}
 independent of the value of $k_1$. The value of $Q^r(\mu,t)$ increases as we move away from the equilateral configuration. The corresponding triangle deformation corresponds to increasing $\chi$  (Figure \ref{fig:triangle}) from $60^{\circ}$ to $180^{\circ}$. Equivalently, the value of $Q^r(\mu,t)$  increases as we  change the shape from an acute triangle to an  obtuse triangle. The right boundary $(\mu=1)$, where we have  the largest values of $Q^r(\mu,t)$,
 corresponds to linear triangles where the three sides of the triangle  are aligned.  
 The bottom right  corner $(\mu,t)=(1,1/2)$ corresponds to  stretched  triangles, and  the top right  corner $(\mu,t)=(1,1)$ corresponds to  squeezed  triangles.  We see that the maximum value of $Q^r(\mu,t)$ occurs along the $\mu=1$ line close to the squeezed limit.  We note that  the  squeezed triangle { where $\mid \kc \mid \rightarrow 0$} is not straight forward  to interpret. {  In the present analysis we have restricted $\kc$ to values  which can be probed using a finite observational volume, and the value $\mid \kc \mid=0$ where the bispectrum is zero is excluded. The entire discussion of squeezed triangles in this paper is restricted to 
  $\mid \kc \mid \rightarrow k_{min}=5 \times 10^{-4} \, {\rm Mpc}^{-1}$ which is comparable to the horizon scale, and   the interpretation is not straightforward } 
 as the value of the bispectrum  depends on how the triangle is squeezed. If we squeeze a linear triangle by first setting  $\mu=1$ and then by taking  the limit $t \rightarrow 1$, then the values of $F_{23}$ and $F_{31}$ diverge (eq.~\ref{eq:b3}), however the diverging parts of these two quantities cancel out to yield a finite value for the bispectrum  (eq.~\ref{eq:b1}). In contrast, the values of $F_{12},F_{23}$ and $F_{31}$ are all  finite if we squeeze an  
  isosceles triangle where  we first set $t=1$  and then take the limit $\mu \rightarrow 1$.
  However, the two above mentioned calculations yield different values for the bispectrum. We therefore conclude that the bispectrum does not have a unique value in the squeezed limit,  rather the result depends on how one approaches this limit.

The right panel of Figure \ref{fig:Q_r} shows thee variation  of  $Q^r(\mu,t)$ along the three boundaries of the allowed region of the $(\mu,t)$ parameter space.   The $x$-axis  corresponds to  $\mu$ or $t$  depending on  the boundary which is being referred to.  The bottom  dashed   curve shows  $Q^r(\mu,1)$ which corresponds to L-isosceles triangles (top boundary of right panel). The left extremity of this curve ($\mu=1/2$) corresponds to equilateral triangles whereas the right extremity ($\mu=1$) corresponds to squeezed triangles. The middle dash-dotted  curve shows  $Q^r(\mu,t)$  as a function of $\mu$ along the boundary $2 \mu t =1$ which corresponds to S-isosceles triangles. Here also  the left extremity ($\mu=1/2$) corresponds to equilateral triangles whereas now the right extremity $(\mu=1)$ corresponds to stretched triangles. The top solid curve shows $Q^r(1,t)$  which 
corresponds to linear triangles (right boundary). Here the left extremity $(t=1/2)$ corresponds to stretched triangles whereas the right extremity $(t=1)$ corresponds to squeezed triangles. The right extremity of the dash-dotted curve and the left extremity of the solid curve both correspond to stretched triangles $(1,1/2)$, and  they both correspond to the same value of $Q^r(\mu,t)$. The right extremity of the solid curve and the dashed curve both correspond to the squeezed limit. However,  as noted earlier the value of the bispectrum depends on how we approach the squeezed limit and we see that the limiting values are different along these two curves. { This is highlighted in the inset of  the right panel which provides a zoom-in view of the squeezed limit. }

As mentioned earlier, in the right panel also we see  that $Q^r(\mu,t)$  is minimum for the equilateral triangle for which $\mu=1/2$.  As $\mu$ is increased the three sides of the triangle get increasingly more aligned (Figure \ref{fig:triangle}) and the value of $Q^r(\mu,t)$ increases with the largest values occurring for the linear triangle when the three sides are parallel.  We find that the value of $Q^r(\mu,t)$  increases by a factor of $\sim 34$  as the equilateral triangle is deformed  to a linear triangle $(\mu=1)$ along either of the two lower curves.
  We find the largest values of $Q^r(\mu,t)$ along the topmost curve which shows this as a function of $t$ with $\mu=1$. 
Considering the variation along the different linear triangles, we find that  $Q^r(\mu,t)$  is nearly constant for $0.5 \le t \le 0.6$, increases for larger $t$ and has a   maxima  at $t \approx 0.94$ beyond which its value  falls. This maxima  is close to the squeezed limit where $k_1 \approx k_2$ and $k_3=(1-t) \, k_1$. We find that the maximum value of  $Q^r(\mu,t)$ occurs at $k_3=k_{eq}$ (matter-radiation equality) which corresponds to the peak of the $\Lambda$CDM power spectrum.   We have tested this by plotting $Q^r(\mu,t)$ for other values of $k_1$ (not shown here). 
The maximum value of $Q^r(\mu,t)$ is found to be approximately $100$ times larger than the minimum value which occurs for equilateral triangles. 

\section{Induced redshift space Bispectrum }
\label{sec:bispec}
The induced bispectrum in redshift space from second order perturbation theory  \citep{Verde:1998zr,Scoccimarro:1999ed}
is given by 
\begin{dmath}\label{eq:bispec}
B^s(\ka,\kb,\kc) = 2  \bl^{-1}(1+\bto\mu_1^2)(1+\bto\mu_2^2)\Big\{ F_2(\ka,\kb)+{\gn\over 2}
+ \mu_3^2\bto G_2(\ka,\kb)-\bl \bto \mu_3 k_3 \Big[ {\mu_1\over k_1}(1+\bto\mu_2^2)
+ {\mu_2\over k_2}(1+\bto\mu_1^2)\Big]\Big\} P(k_1) P(k_2) + 
{\rm cyc ...}\, ,
\label{eq:zb1}
\end{dmath}
where $\bto$  is the linear redshift distortion parameter and  
\begin{equation}
G_2(k_1,k_2)={3\over 7}+{\ka\cdot \kb\over 2}\left({1\over k_1^2}+{1\over k_2^2}\right)+{4\over7}{(\ka \cdot \kb)^2\over k_1^2 k_2^2},
\label{eq:G2}
\end{equation}
refers to the second order kernel for the divergence of the peculiar velocity 
\citep{Goroff:1986ep}. Here we find it useful to use  
\begin{equation}
F_2(\ka,\kb) = G_2(\ka,\kb) + \Delta G(\ka,\kb)  
\label{eq:F2}
\end{equation}
where 
\begin{equation}
\Delta G(\ka,\kb)= {2\over 7}\left[1-{(\ka \cdot \kb)^2\over k_1^2k_2^2}\right]\,.
\label{eq:dG2}
\end{equation}
The various terms $G_2(\ka,\kb)$ and $\Delta G(\ka,\kb)$ 
 are dimensionless functions which depend on the shape of the triangle $(\ka,\kb,\kc)$, and these have very simple algebraic expressions in terms of $(\mu,t)$. Using the compact notation $G_{12}(\mu,t)=G_2(\ka,\kb)$ and $\Delta G_{12}(\mu,t)=\Delta G(\ka,\kb)$, etc.  we can express these as 
\begin{eqnarray}
G_{12}(\mu,t) &=& \frac{1}{14} \left(8 \mu ^2-7 \mu  t-\frac{7 \mu }{t}+6\right)\nn\\
G_{23}(\mu,t)&=&-\frac{ 6 \mu ^2 t+t-7 \mu}{14 s^2 t}\nn\\
G_{31}(\mu,t)&=&\frac{t^2 \left(-6 \mu ^2+7 \mu  t-1\right)}{14 s^2}
\end{eqnarray}
and 
\begin{equation}
s^2 t^{-2} \Delta G_{31} = \, s^2 \, \Delta G_{23}= \Delta G_{12}=\frac{2}{7} \left(1-\mu ^2\right) \,. 
\end{equation}
We note that all the $\Delta G$s are zero for linear triangles $(\mu=1)$.

For calculating the various angular  moments of  the redshift space bispectrum we express  it  as 
\begin{dmath} 
B^s(\ph,k_1,\mu,t)  = 2 \bl^{-1} \Big\lbrace R G_{12} + \S{12} \left[\Delta G_{12}+{\gn\over 2}\right]
-\bl \T{12} \Big\rbrace  P^r(k_1) \, P^r(k_2)+{\rm cyc ...}
\label{eq:zb3}
\end{dmath}
Here the anisotropy or orientation dependence of the redshift space bispectrum is completely quantified by the functions  
\begin{equation}
R(\ph,\bto,\mu,t) = (1+\bto\mu_1^2)(1+\bto\mu_2^2)(1+\bto\mu_3^2)   \,, 
\label{eq:zb4}
\end{equation}
\begin{equation}
\S{12}(\ph,\bto,\mu,t)=(1+\bto\mu_1^2)(1+\bto\mu_2^2)  
\label{eq:zb5}
\end{equation}
and 
\begin{dmath}
\T{12}(\ph,\bto,\mu,t) = {\bto\over 2}(1+\bto\mu_1^2)(1+\bto\mu_2^2)\mu_3k_3\times
\Big[{\mu_1\over k_1}(1+\bto\mu_2^2)+{\mu_2\over k_2}(1+\bto\mu_1^2)\Big] \,.
\label{eq:zb6}
\end{dmath}
The angular multipoles of the redshift space bispectrum can be expressed in terms of the angular multipoles of $R, \, \S{12}, \, \T{12}$, etc as    
\begin{eqnarray}
\bar{B}^{m}_{\ell}(k_1,\mu,t)=
 &=& 2 \bl^{-1} \Big\lbrace \RB^m_{\ell}\,  G_{12} + \SB{m}{\ell}{12}  \left[\Delta G_{12}+{\gn\over 2}\right]\nn\\&&
-\bl \TB{m}{\ell}{12}  \Big\rbrace  P^r(k_1) \, P^r(k_2)+{\rm cyc ...}
\label{eq:zb7}
\end{eqnarray}
where 
\begin{eqnarray}
\RB_\ell^m(\bto,\mu,t)=\sqrt{\frac{(2 \ell +1)}{4 \pi}} \int [Y^m_{\ell}(\ph)]^* R(\ph,\bto,\mu,t) \, d\Omega_{\ph} \, .
\label{eq:zb8}
\end{eqnarray}
refers to the   angular angular multipoles of $ R(\ph,\bto,\mu,t)$, 
and the other multipole moments $\SB{m}{\ell}{12}, \TB{m}{\ell}{12}$ etc. have been  defined similarly. 
The different multipole moments $\RB^m_{\ell}, \SB{m}{\ell}{12}, \TB{m}{\ell}{12} $ etc. 
are all  polynomials of the form $c_0  + c_1 \beta_1 + c_2 \bto^2 ...$ where  the smallest power of $\bto$  is  $\ell/2$ and  the largest power of $\bto$ is $3,2$ and $4$ for   $\RB^{m}_{\ell}, \SB{m}{\ell}{12}$ and 
$\TB{m}{\ell}{12}$ respectively. The coefficients of the polynomials depend on $\ell,m$ and the shape of the triangle $(\mu,t)$.

$R(\ph,\bto,\mu,t)$  (eq.~\ref{eq:zb4}) corresponds to the redshift space enhancement of the bispectrum that arises if each of  
the three terms $\Delta(\ka)$, $ \Delta(\kb)$ and  $\Delta(\kc)$ in equation~(\ref{eq:a1}) is subjected  to  linear RSD.  The angular multipoles  $\RB_\ell^m(\bto,\mu,t)$ have non-zero values for even $\ell$ in the range $\ell \le 6$, while $m \le \ell$ with $m$ being restricted to $m \le 4$. 
The angular multipoles  $\RB_\ell^m(\bto,\mu,t)$ have been studied in detail in \citetalias{Bharadwaj:2020wkc}, and when required we  use these  results here.  

Considering the $S$ terms (eq.~\ref{eq:zb5}), the angular multipoles   have non-zero values for even $\ell$ in the range $\ell \le 4$, while $m \le \ell$. We see  that for linear triangles $(\mu=1)$ the three $S$ terms  all reduce to the same value  
\begin{eqnarray}
\S{12} = \S{23} =\S{31} = (1+\bto\mu_1^2)^2\, .
\label{eq:zb9}
\end{eqnarray}
which is exactly the enhancement factor of the power spectrum due to linear RSD. The multipole moments of this enhancement factor have been  extensively  studied  \citep{Hamilton:1997zq}, and denoting these as $A_{\ell}$ we have  
$A_0= 1+ 2 \bto/3  + \bto^2 /5$, $ A_2=4(\bto/3  + \bto^2/7)$  and  
$A_4=8 \bto^2 /35$. We find that we can express all the multipole moments with $m=0$ in the form    
\begin{dmath}
\SB{0}{\ell}{12}(\bto,\mu,t) =A_\ell(\bto) + \Delta G_{12}(\mu)  \,[C_{12}]_{\ell}(\bto,\mu,t) \, ,
\label{eq:zb10}
\end{dmath}
where the second term in the R.H.S. is zero for linear triangles. Considering the $S$ terms with $m \neq 0$, these are all zero for linear triangles $(\mu=1)$, and we find that we find it convenient to express  these in the form 
\begin{equation}
\SB{m}{\ell}{12}(\bto,\mu,t)=(1-\mu^2)^{m/2} \, [D_{12}]_\ell^m(\bto,\mu,t) \, ,
\label{eq:zb11}
\end{equation}
The expressions  needed to calculate the  multipole moments of the $S$ terms  $(A_{\ell}, [C_{12}]_\ell, [D_{12}]_\ell$, etc.) are  presented in Appendix \ref{appendix0} instead of  the main body of the text.

Considering the $T$ terms (eq.~\ref{eq:zb6}), the angular multipoles   have non-zero values for even $\ell$ in the range $\ell \le 8$, while $m \le \ell$ with $m \le 6$. The expressions for the multipole moments here  are rather  lengthy and we have presented these in Appendix \ref{appendix1} instead of  the main body of the text. 

We have used the multipole moments of $R$, $S$ and $T$ in eq.~(\ref{eq:zb7}) to  
calculate $\bar{B}^{m}_{\ell}(k_1,\mu,t)$  the  multipole moments of the redshift space bispectrum.
\footnote{Python scripts for calculating these moments are available at 
\href{https://github.com/arindam-mazumdar/rsd-bispec}{\tt https://github.com/arindam-mazumdar/rsd-bispec}.}

\section{Results}\label{sec:results}

Here we analyze  all the non-zero  multipole moments of the redshift space bispectrum. For this purpose we shall consider $Q^{m}_{\ell}(k_1,\mu,t)$ which is defined exactly identically as  $Q^r(k_1,\mu,t)$ (eq.~\ref{eq:b4})  except that we have replaced $B^r(k_1,\mu,t)$ with $\bar{B}^{m}_{\ell}(k_1,\mu,t)$. Unless mentioned otherwise, we shall focus on $Q^m_{\ell}(\mu,t)$ which shows the results  for $k_1=0.2 \, {\rm Mpc}^{-1}$ and the reference model where $\bto=1, \bl=1$ and $\gn=0$.  Table~\ref{tab:1}  shows the maximum and minimum values of all the non-zero $Q^m_{\ell}(\mu,t)$ along with the $(\mu,t)$ values where these occur. 

\begin{figure}
 \includegraphics[width=0.5\textwidth]{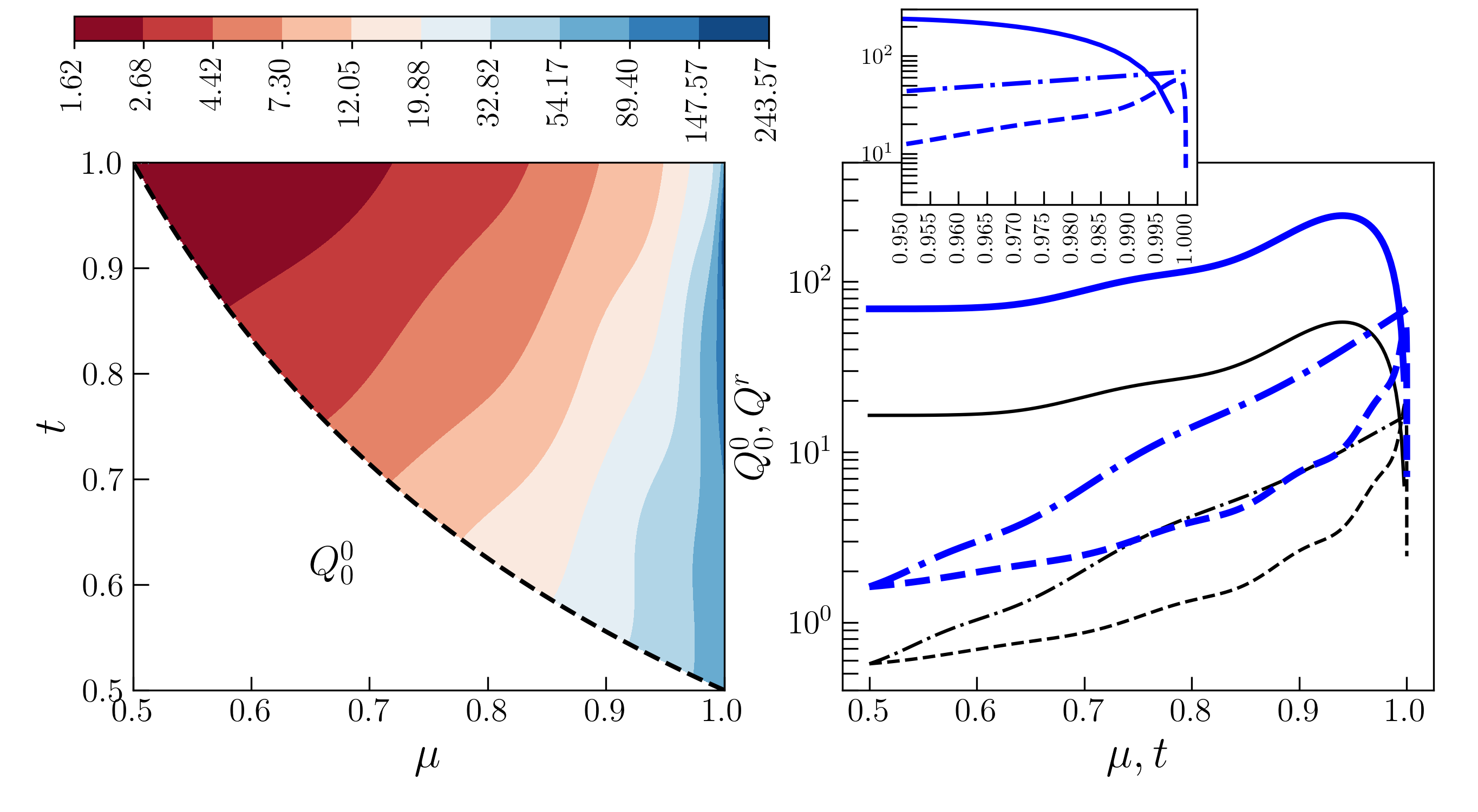}
\caption{Same as Figure \ref{fig:Q_r} except that this shows $Q^0_0(\mu,t)$. The right panel shows 
$Q^r(\mu,t)$  (black) in addition to $Q^0_0(\mu,t)$ (blue).}
\label{fig:mono1}
\end{figure}

 The left panel of  Figure \ref{fig:mono1} shows how  the dimensionless  monopole bispectrum $Q^0_0(\mu,t)$  varies with the shape of the triangle.  We see that the pattern is very similar to that seen 
in the left panel  of Figure \ref{fig:Q_r} for the real space bispectrum $Q^r(\mu,t)$. Here also the value is minimum for equilateral triangles, 
and we have the largest values for linear triangles. For equilateral triangles we have
\begin{eqnarray}
Q_0^0(1/2,1) &= & \frac{\bto ^3}{490}+\frac{\bto ^2 \gn }{10}+\frac{3 \bto ^2}{35}-\frac{\bl \bto ^4}{315}+\frac{\bl \bto ^2}{5}+\frac{\bl  \bto }{3}\nn\\ &&+\frac{2 \bto  \gn }{3}+\frac{3 \bto }{7}+\gn +\frac{4}{7} \, 
\label{eq:c2}
\end{eqnarray}
which is independent of $k_1$. This goes over to $Q^r(1/2,1)$ (eq.~\ref{eq:b5})  for $\beta=0$ . Comparing  eq.~(\ref{eq:c2}) with eq.~(\ref{eq:zb7}) we see that the terms involving $\gn$ arise from the $S$ terms, the terms involving $\bl$ arise from the $T$ terms and the remaining terms  are a combination of the contributions from the $R$ and $S$ terms.  We have checked that our expression  for the monopole of the  redshift space bispectrum matches   the  results presented  in \citet{scoccimarro98-NL-bispec} for the   equilateral triangle. A more general comparison is not possible as the parameterization of the triangle shape and size dependence is quite different. 

The right panel of 
Figure \ref{fig:mono1} is identical to the right panel of Figure \ref{fig:Q_r} except that this shows  the monopole $Q^0_0(\mu,t)$ (blue), the real space bispectrum  $Q^r(\mu,t)$  (black) has also been shown for comparison.  We see that the two sets of  curves are very similar except for the fact that the  values of  $Q^0_0(\mu,t)$ are larger than those of $Q^r(\mu,t)$. The enhancement due to RSD is found to be $\sim 2.8$ for equilateral triangles. Further, we find that the enhancement is munimum for equilateral triangles and it increases as the shape is deformed towards linear triangles. We find that the enhancement is $\sim 4.2$ for linear triangles irrespective of the value of $t$. The location of the maximum value of  $Q^0_0(\mu,t)$  coincides with that of $Q^r(\mu,t)$. As discussed earlier, this occurs at $\mu=1, t \approx 0.95$ which corresponds to $k_1 \approx k_2$ and $k_3 \approx k_{eq}$.  
Like the real space bispectrum, we find that the squeezed limit is not uniquely defined for the monopole of the redshift space bispectrum and the result depends on how we approach the squeezed limit.  { This is highlighted in the inset of  the right panel which provides a zoom-in view of the squeezed limit. }

\begin{figure*}
     \includegraphics[width=0.9\textwidth]{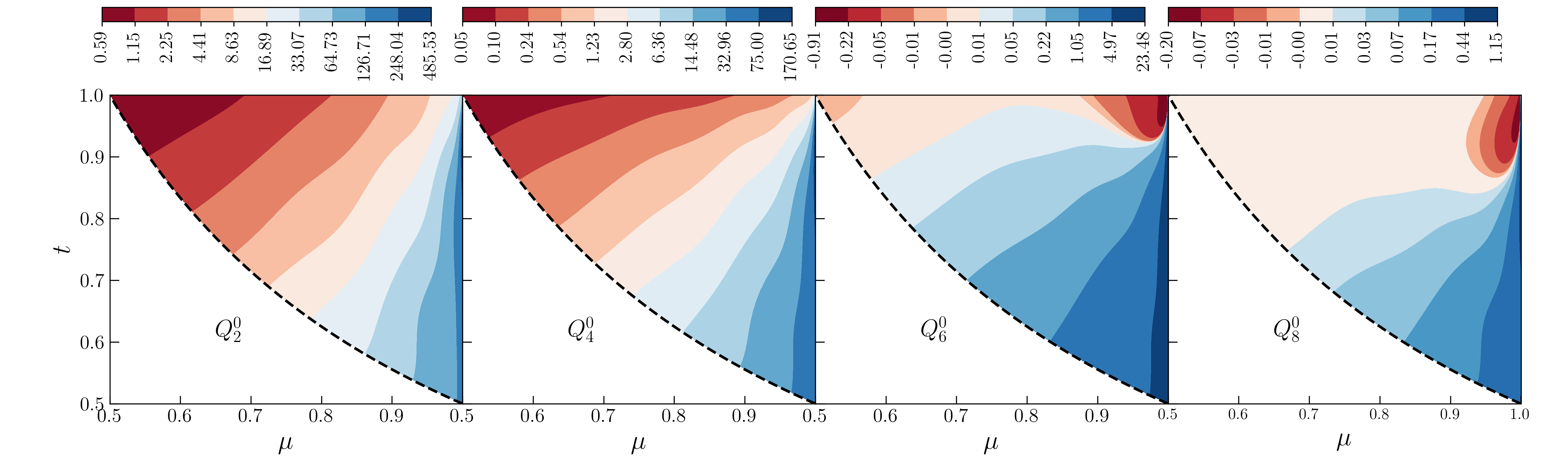}
    \caption{Shape dependence of all the $\ell >0$ non-zero moments $Q_{\ell}^0(\mu,t)$.  }
    \label{fig:m_0} 
\end{figure*}
   
\begin{figure*}
    \centering
    \includegraphics[width=0.9\textwidth]{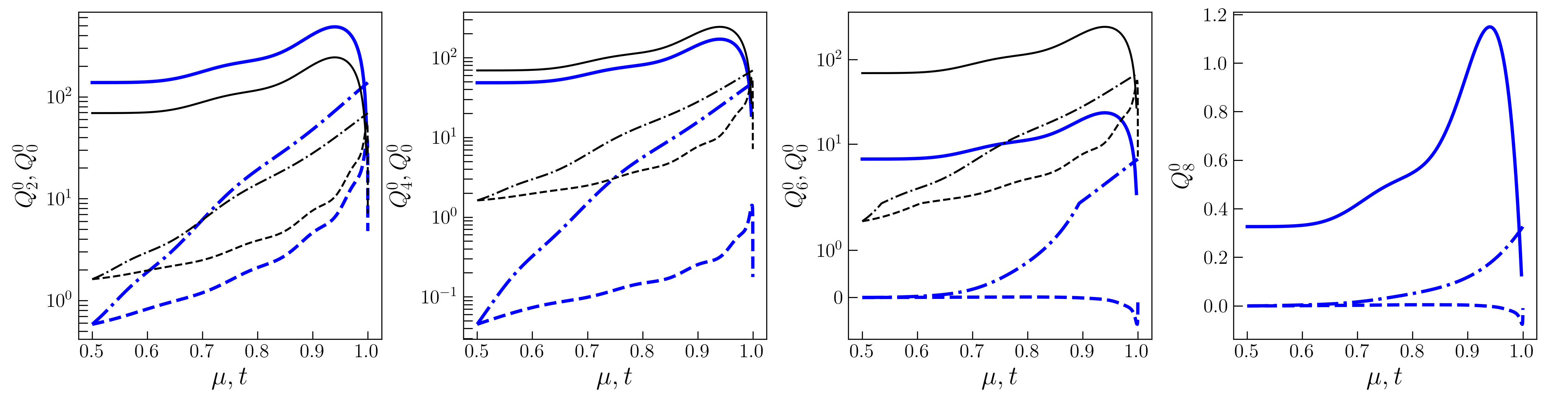}
    \caption{Each panel here shows $Q_{\ell}^0$ (blue)  along the boundaries of the allowed $(\mu,t)$ region in the corresponding panels of Figure \ref{fig:m_0},  $Q_0^0$ (black) is also shown for comparison. }
    \label{fig:m_01D}
\end{figure*}

 Figure \ref{fig:m_0} shows the $m=0$ component of the quadrupole $(\ell=2)$ and all the higher multipoles  $(\ell=4,6,8)$ which are predicted to have non-zero values. We find that for $\ell=2$ and $4$ the  dependence on the shape of the triangle is very similar to that seen in Figures \ref{fig:Q_r}  and \ref{fig:mono1}  for  $Q^r(\mu,t)$  and  $Q^0_0(\mu,t)$ respectively.   We have the minimum 
 value for the equilateral triangle, and the value increases as we move towards linear triangles where we  have the largest values. Considering $\ell=6$ and $8$,  here also  the pattern is similar to that at lower  $\ell$. We have very small values of $Q^0_{\ell}$ 
 in the vicinity of  equilateral triangles,    and the value increases as we move towards obtuse linear triangles where we have the largest values. The difference, however, is that unlike the lower multipoles we find a region of negative values near  the top boundary $(t=1)$. The magnitude of these negative values increases if we approach the squeezed limit along the top boundary, 
 this being particularly pronounced for $\ell=8$. 
 
Figure \ref{fig:m_01D} shows the value of $Q^0_{\ell}(\mu,t)$ along the boundaries of the allowed $(\mu,t)$ region plotted in the different panels of  Figure \ref{fig:m_0}.  For $\ell=2,4$ and $6$  the monopole  $Q^0_{\ell}(\mu,t)$ is also shown in the respective panels for comparison.  We see that the values  of $Q^{0}_{\ell}$ get  smaller as $\ell$ is increased (Table~\ref{tab:1}), and for $\ell=8$ the values are so small 
that it does not make sense to show them  together  with $Q^0_0$.  We see that the solid curves  which  show $Q^0_{\ell}(1,t)$ for  linear triangles  are  very similar for all the $\ell$ values (including $\ell=0$) shown here.
The same is also true for the  dashed-dot curves  which show $Q^0_{\ell}(\mu,1/(2 \mu))$ for the S-isosceles triangles. 
However,  this is not true for the dashed curves which show $Q^0_{\ell}(\mu,1)$ for the L-isosceles triangles. Here 
$\ell=0,2,$ and $4$ all show a similar behaviour   while $6$ and $8$ show a different behaviour where $Q^0_{\ell}(\mu,1)$ has negative values. 
The various  multipole moments with $\ell >0$ all quantify the anisotropy or orientation dependence which arises due to RSD, and the ratio $Q^m_{\ell}/Q_0^0$ provides a quantitative  estimate  of the relative strength of this anisotropy. Considering $m=0, \ell=2$  we find 
that the relative strength of the anisotropy is minimum for equilateral triangles where we have $Q^0_2(1/2,1) \approx 0.36 \, Q^0_0(1/2,1)$. The low level of anistropy for equilateral triangles here may be attributed to the fact that the three vectors $\ka,\kb$ and  $\kc$ are differently oriented relative to each  other. When one of them is aligned with  $\zh$ for which  the RSD is maximum, the other two vector  are  inclined  with respect to $\zh$. We see that the relative strength of the anisotropy increases as the shape is deformed towards linear triangles where the three vectors $\ka,\kb$ and  $\kc$ are aligned. We have the largest anisotropy  $(Q^0_2(1,t) \approx 2 \, Q^0_0(1,t))$  for linear triangles irrespective of the value of $t$. Interestingly the quadrupole is larger than the monopole $(Q^0_2 > Q_0^0)$ over a significant portion of the $(\mu,t)$ parameter space. As discussed earlier, the maxima of $Q^0_2(\mu,t)$ 
occurs at $\mu=1, t \approx 0.94$ which corresponds to $k_1 \approx k_2$ and $k_3 \approx k_{eq}$.  
The position of the maxima coincides with that for $Q^0_0(\mu,t)$ and $Q^r(\mu,t)$. 
Like the real space bispectrum, we find that the squeezed limit is not uniquely defined for the quadrupole of the redshift space bispectrum and the result depends on how we approach the squeezed limit.  $Q^0_4$ is very similar to 
$Q^0_2$  except that the values are smaller with $Q^0_4(1/2,1) \approx 0.028 \, Q^0_0(1/2,1)$ and $Q^0_4(1,t) \approx 0.7 \, Q^0_0(1,t)$ for equilateral and linear triangles respectively.  Considering $\ell=6$ we find that the values are even smaller with  $Q^0_6(1,t)\approx 0.096 \, Q^0_0(1,t)$ for linear triangles. The value is negative for equilateral triangles $Q^0_6(1/2,1) \approx -0.0019 \, Q^0_0(1/2,1)$, the negative values  continue along $t=1$ and increases in magnitude as we approach the squeezed limit. However note that the squeezed limit continues to have a positive value if we approach  it along $\mu=1$. For $\ell=8$ the results are very similar to $\ell=6$ except that the values are even smaller with $Q^0_8(1/2,1) \approx -0.00035 \, Q^0_0(1/2,1)$ and $Q^0_8(1,t)\approx 0.0047 \, Q^0_0(1,t)$
for equilateral and linear triangles respectively.

Considering equilateral triangles, we obtain relatively compact  expressions for the various $m=0$ multipole moments which we present below
\begin{eqnarray}
Q_2^0(1/2,1) &=&\frac{\bto ^3}{588}+\frac{\bto ^2 \gn }{14}+\frac{3 \bto^2}{49}-\frac{2}{693} b_1 \bto ^4+\frac{b_1 \bto ^2}{7}+
\nn\\&&
\frac{b_1 \bto}{6}+\frac{\bto  \gn }{3}+\frac{3 \bto }{14} \,,
\label{eq:d3}
\end{eqnarray}

\begin{dmath}
Q_4^0(1/2,1) =\frac{27 \bto ^3}{43120}+\frac{9 \bto ^2 \gn }{560}+\frac{27 \bto ^2}{1960}-\frac{27 b \bto ^4}{20020}+\frac{9 b \bto ^2}{280} \,,
\end{dmath}
\begin{equation}
Q_6^0(1/2,1)= \frac{59 \bto ^3}{12936}-\frac{1697 b \bto ^4}{221760}   \,,  
\end{equation}
\begin{equation}
Q_6^0(1/2,1)= -\frac{29 b \bto ^4}{51480} \,.
\end{equation}
 We have compared our expression  for the $m=0$  quadrupole moment  with  the  results presented  in \citet{scoccimarro98-NL-bispec} for the   equilateral triangle.  We find that the two  results match  except for 
the terms involving $\gn$ for which  our results have  twice the value.   As mentioned earlier, the terms involving $\gn$ arise from the $S$ terms (eq.~\ref{eq:zb7}). The same $S$ terms also contribute to the terms which do not have $\gn$ or $\bl$, these however match the results in  \citet{scoccimarro98-NL-bispec}. The cause of this discrepancy is not clear at present. It may however be noted that the different $m$ components of the multipole moments $\bar{B}_{\ell}^m(k_1,\mu,t)$ are not uniquely defined, and these can vary (through a rotation matrix) depending  on the choice of $x,y,z$ axis. 


 \begin{figure*}
     \includegraphics[width=0.9\textwidth]{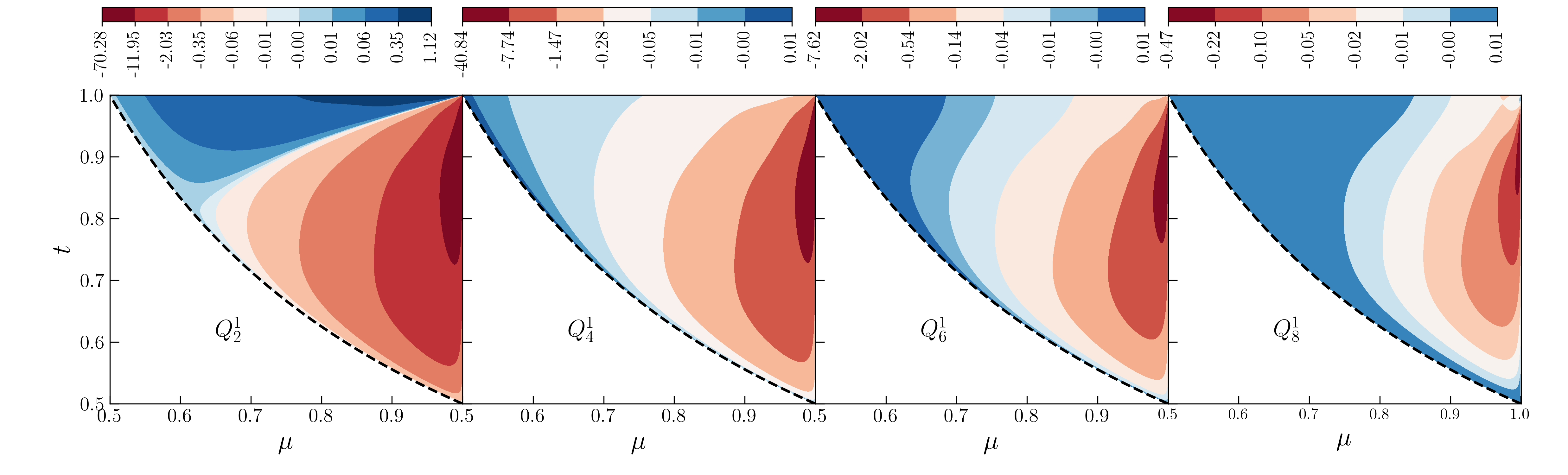}
    \caption{Shape dependence of all the  non-zero moments $Q_{\ell}^1(\mu,t)$.} 
    \label{fig:m_1} 
\end{figure*}

\begin{figure*}
     \includegraphics[width=0.9\textwidth]{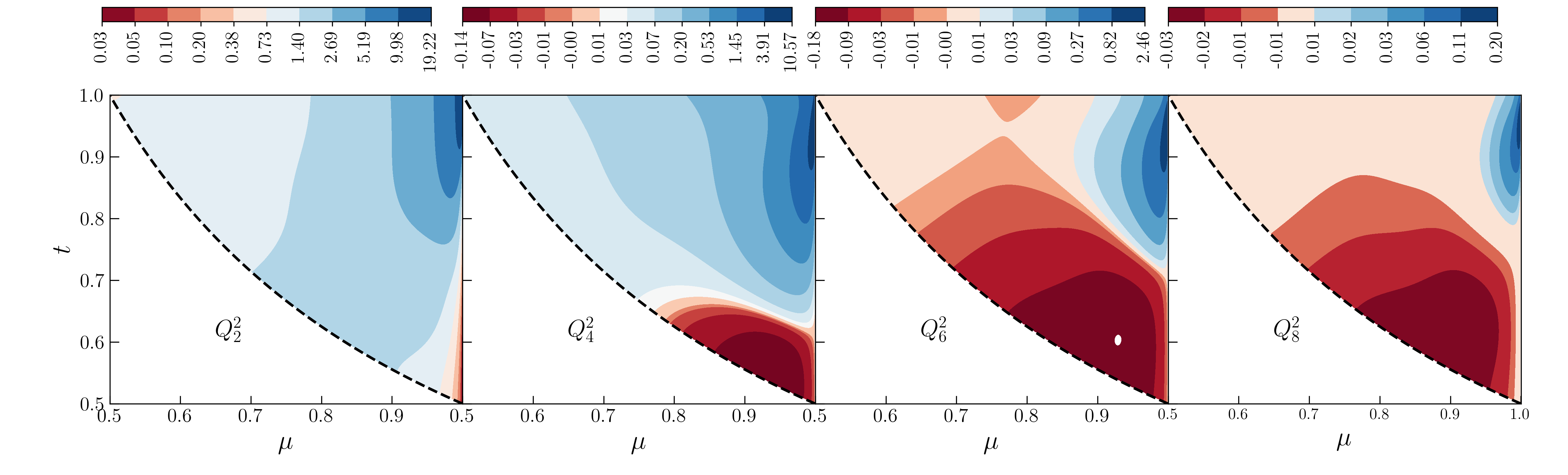}
    \caption{Shape dependence of all the  non-zero moments $Q_{\ell}^2(\mu,t)$. }
    \label{fig:m_2} 
\end{figure*}

 Figures \ref{fig:m_1} and \ref{fig:m_2} show all the non-zero 
 $Q^m_{\ell}(\mu,t)$  for $m=1$ and $2$ respectively.  We see that the patterns  are   quite different  from  those  seen in Figures \ref{fig:Q_r}, \ref{fig:mono1} and \ref{fig:m_0} for  $Q^r(\mu,t)$,  $Q^0_0(\mu,t)$ and $Q^0_{\ell}(\mu,t)$
respectively. Considering $m=1$, the results are similar for all $\ell$ values. We see  that  
 $Q^1_{\ell}(\mu,t)$ is zero for equilateral triangles,  and also along the lower  boundary (S-isosceles triangles)
and the right boundary (linear triangles).  We have positive values in the upper left region  around  the equilateral triangle, the positive region extends along the top of the figure all the way to the squeezed limit. For $\ell=2$ we encounter the  maximum value of $Q^1_2 \, (\approx 1)$   at the squeezed limit if we approach it  along $t=1$.  For higher $\ell$, the maximum value is close to the equilibrium triangle and the value gets smaller as $\ell$ increases.  
$Q^1_{\ell}(\mu,t)$ has negative values for obtuse triangles $\mu >t$, and to some extent  the negative regions extends across the $\mu=t$ line  into the acute triangles also.  
The minimum value of $Q^1_2(\mu,t) \, (\approx -70)$  occurs close to the squeezed limit very near the $\mu=1$  boundary ( at $t=0.95$),  beyond the minima the value 
of $Q^1_2(\mu,t)$  sharply falls to zero  at $\mu=1$. 
The other $\ell$ values show a similar behaviour, however the magnitude of the minimum value is a factor of $1.7,9.7$ and $155.27$ smaller compared to $\ell=2$ for $\ell=4,6$ and $8$ respectively.  Considering $m=2$ (Figure \ref{fig:m_2}),  for all values of $\ell$  we find  that $Q^2_{\ell}(\mu,t)$ has positive values near the squeezed limit with a maxima very close to the squeezed limit beyond which it sharply falls to zero at $\mu=1$.  $Q^2_2(\mu,t)$ is positive everywhere,  the maximum value   $\approx 19$  occurs approximately at $t=1, \mu = 0.98$ beyond which it sharply falls to zero at $\mu=1$. The location of the maxima is the same for other $\ell$, however the value falls by a factor of $1.84,7.93$ and $95.24$ compared to $\ell=2$ for $\ell=4,6$ and $8$ respectively.  We see that for all $\ell$ the minima is near the stretched limit where the values are  negative for $\ell > 2$. We see that $Q^2_{\ell}(\mu,t)$ has a  negative value over much of the $(\mu,t)$ space for $\ell=6$ and $8$.

\begin{figure}
\includegraphics[width=0.45\textwidth]{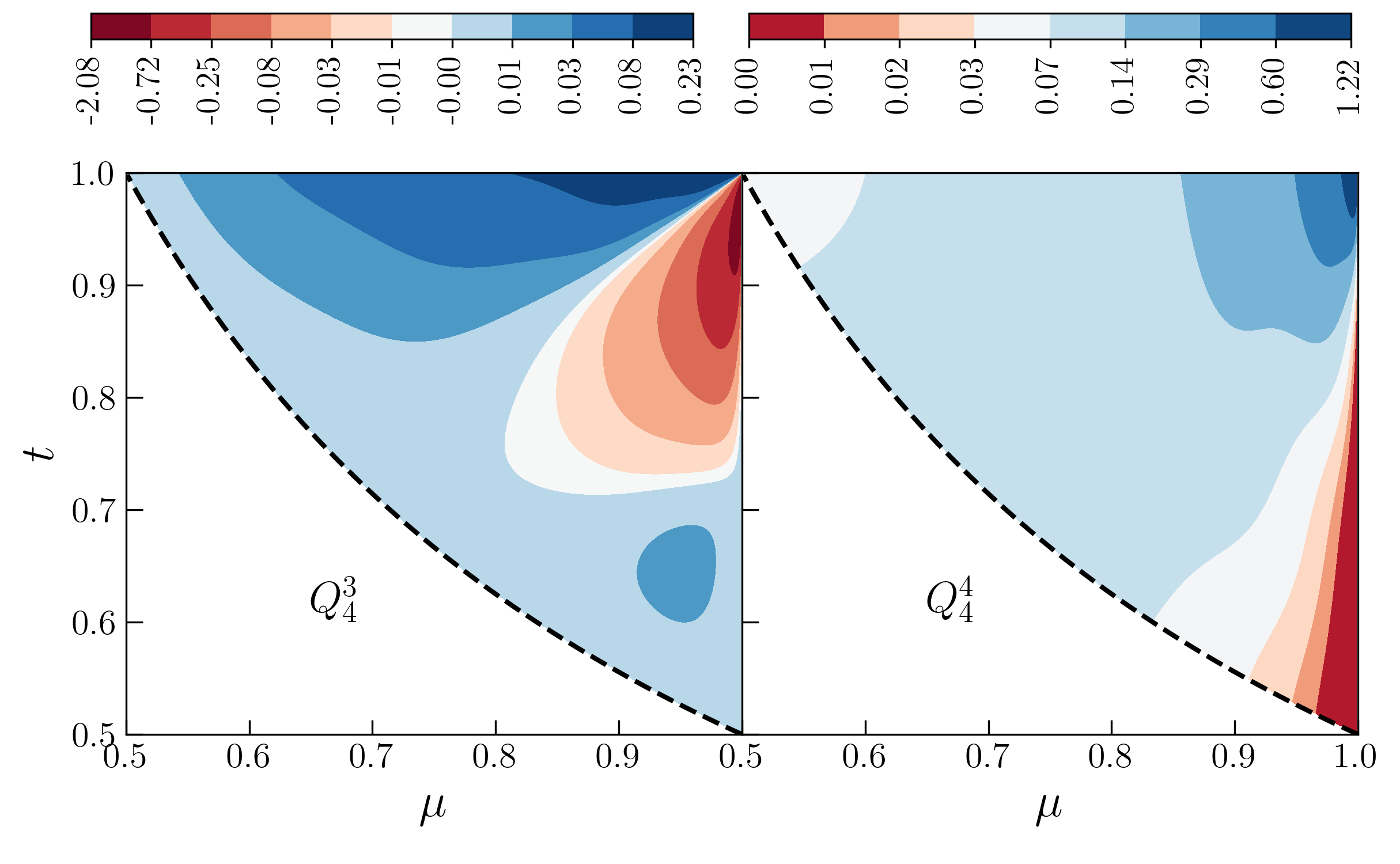}
    \caption{Shape dependence of  $Q_{4}^3(\mu,t)$ and $Q_{4}^4(\mu,t)$.} 
    \label{fig:l_4} 
\end{figure}

Figure \ref{fig:l_4}  shows $Q^m_{4}(\mu,t)$  for $m=3$ and $4$. We see that like $Q^1_{4}(\mu,t)$, $Q^3_{4}(\mu,t)$ also is zero for equilateral triangles, S-isosceles triangles and linear triangles and it  has a small positive value ($\sim 0.1$ and smaller) over much of the $(\mu,t)$ space. The maximum value $(\sim 0.2)$ occurs near the squeezed limit if we approach it close to  the top boundary $(t \approx 1)$. We find negative values for obtuse triangles $(\mu > t)$ near  the squeezed limit, and the minimum value $(\approx -2.08)$ occurs very close to the squeezed limit({ $\mu=0.999,t=0.976$}).  Considering $Q^4_{4}(\mu,t)$, we see that this has positive values  over all of $(\mu,t)$ space, and is zero for linear triangles. The maximum  value $(\approx 1.21)$ occurs  very close to the squeezed limit. 

\begin{figure*}
     \includegraphics[width=0.9\textwidth]{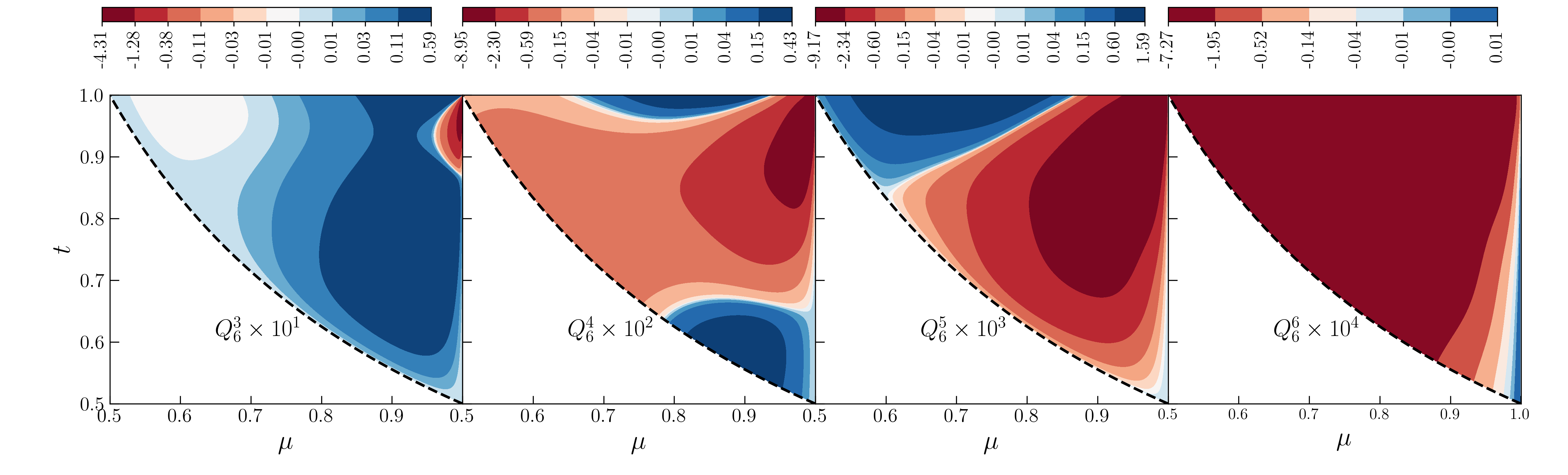}
    \caption{Shape dependence of all the $m >2$ non-zero moments $Q_{6}^m(\mu,t)$.}
    \label{fig:l_6} 
\end{figure*}

\begin{figure*}
     \includegraphics[width=0.9\textwidth]{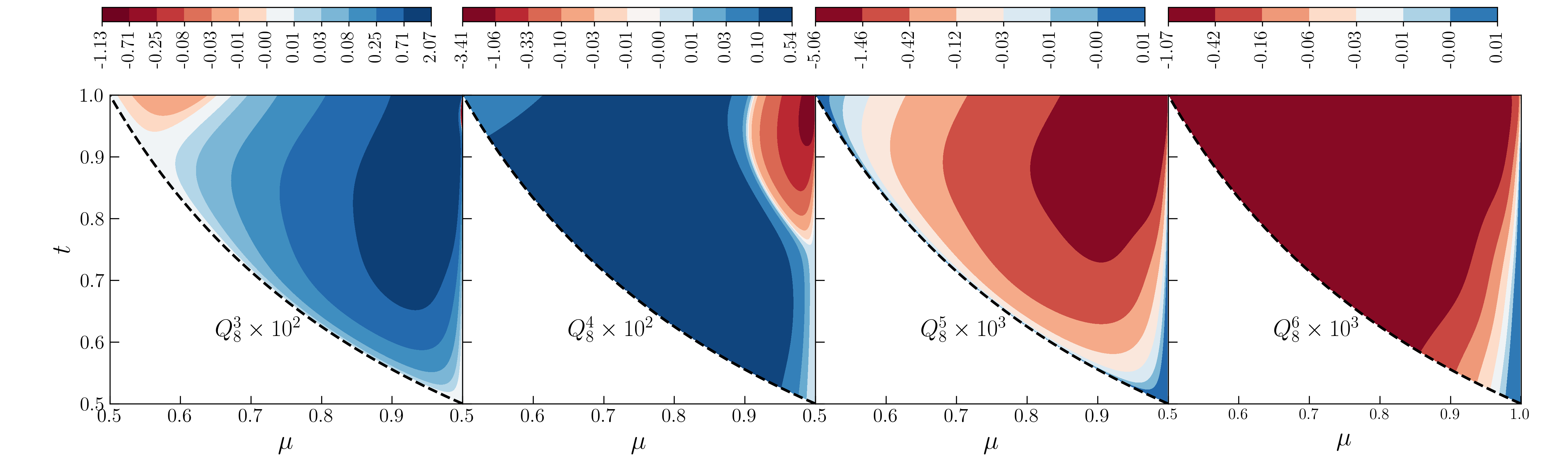}
         \caption{Shape dependence of all the $m >2$ non-zero moments $Q_{8}^m(\mu,t)$.}
    \label{fig:l_8} 
\end{figure*}

For completeness, we have shown all the remaining non-zero  multipoles $(m=3,4,5,6)$  in Figures \ref{fig:l_6} and \ref{fig:l_8} for $\ell=6$ and $8$ respectively. For all the odd $m$ the value is zero for equilateral and  S-isosceles triangles, whereas for linear triangles  the value is zero for both odd and even $m$. The values of  $Q^m_{\ell}(\mu,t)$ become extremely small as $m$ and $\ell$ are increased, and we have $\mid Q^m_{\ell}(\mu,t) \mid < 0.1$ for all the results shown in these two figures except for $Q^3_6$. We see that $ Q^3_{6}(\mu,t) $  has  relatively larger negative values around the  minima  $(\approx -0.43$) which occurs for obtuse triangles very close to the squeezed limit.

\input{table1.tex}
\section{Summary and Discussion}\label{sec:summ}
This paper and \citetalias{Bharadwaj:2020wkc} together present  the formulas needed to calculate all  the multipole moments 
$\bar{B}_{\ell}^m(k_1,\mu,t)$ of the redshift space bispectrum which are predicted to be non-zero at second order perturbation theory.  Equation~(\ref{eq:zb7}) represents our final result where the  $R,S$ and $T$ terms are given  in \citetalias{Bharadwaj:2020wkc}, Appendices~\ref{appendix0} and \ref{appendix1}  respectively.   
In this paper we have analyzed $Q^m_{\ell}(\mu,t)$ which quantifies the dependence on the shape of the triangle keeping the other quantities  which affect $\bar{B}_{\ell}^m(k_1,\mu,t)$ fixed at 
$k_1=0.2 \, {\rm Mpc}^{-1}$, $\bto=1, \bl=1,\gn=0$.  The results are not very different for other values of $k_1$  close to 
$0.2 \, {\rm Mpc}^{-1}$.  We have also analyzed $Q^r(\mu,t)$ which corresponds to the real space bispectrum. 

We find that $Q^r(\mu,t), Q^0_0(\mu,t), Q^0_2(\mu,t)$ and $Q^0_4(\mu,t)$ all show very similar behaviour where the values are positive everywhere, we have the smallest value for equilateral triangles, the value increases as the triangle is deformed towards a linear triangle and we have the largest values for linear triangles.  In all of these cases we have the maximum value at $\mu=1, t \approx 0.94$ which is very close to the squeezed limit and where $k_1 \approx k_2$ with $k_3 \approx k_{eq}$.  Considering any triangle shape $(\mu,t)$ the monopole $Q^0_0$ exceeds $Q^r$, the RSD enhancement being minimum for equilateral and maximum for linear triangles. The quadrupole $Q^0_2$ exceeds $Q^0_0$ for a significant region of $(\mu,t)$ space in the proximity of linear triangles,  however $Q^0_2$ is smaller than $Q^0_0$ around the equilateral configuration. The values of $Q^0_4$ are smaller than those of  $Q^0_0$ for the entire $(\mu,t)$ space, the two differing by a factor of $3$ and $30$ for linear and equilateral triangles respectively. The $m=1$ and $m=2$ multipoles show distinctly different patterns. For $Q^1_2$ we have negative values of large magnitude near the squeezed limit for obtuse triangles, and the maximum magnitude is roughly $4$ times smaller than the maxima of $Q^0_0$. $Q^2_2$ is positive throughout, and has a maxima very close to the squeezed limit with maximum  value $\approx 13$ times  smaller than the maxima of $Q^0_0$. The higher multipoles all show a rich variety of shape dependence, however the values of $\mid Q^m_{\ell} \mid$ falls drastically as $\ell$ and $m$ are increased. 

The various patterns in the shape dependence of $Q^m_{\ell}(\mu,t)$ shown  here are purely a consequence of non-linear gravitational clustering as we have used $\gn=0$ throughout. Non-linear bias $(\gn \neq 0)$, if present, will cause the shape dependence  of $Q^m_{\ell}(\mu,t)$ to differ from that shown here.  
The $k_1$ and shape $(\mu,t)$ dependence of the various multipole moments of the bispectrum contains a wealth of cosmological information. It has been proposed that the shape dependence of the bispectrum can be used to determine the parameters $\bto,\bl$ and $\gn$.  We propose to study these issues in future. 

{Upcoming future galaxy surveys like Euclid \citep{euclid2011} and DESI \citep{desi:2013} are aimed to cover   large volumes of the order of several ${\rm Gpc}^3$. Following   \citet{Scoccimarro:1997st} (and also  \citealt{Tellarini:2016sgp}) 
we have estimated the expected variance of the bispectrum for such surveys. Expressing this in terms of $Q$ we obtain 
$\Delta Q \approx 0.001$ for equilateral triangles with $k_1= 0.2 \, {\rm Mpc}^{-1}$ and  bin width  $(\delta k_1)= k_1/10$.  These estimates indicate that in future it may be possible to observe a large number of the bispectrum multipoles considered  here 
(Table~\ref{tab:1}) all the way to $Q^4_6$ or $Q^5_6$ . In subsequent works we plan to present more quantitative predictions for observing these multipoles. }  

\section*{Acknowledgements}
DS acknowledges support from the Azrieli Foundation for his Postdoctoral Fellowship.

\section*{Data availability} The data underlying this article were generated by the sources available in public domain
{\tt https://lesgourg.github.io/class\_public/class.html} and
{\tt https://github.com/arindam-mazumdar/rsd-bispec}

\bibliographystyle{mnras}
\bibliography{reference}

\appendix
\section{Multipole moments of the  $S$ terms}\label{sec:a1}
\label{appendix0}
\input{Sterms.tex}

\section{Multipole moments of the  $T$ terms}\label{sec:a2}
\label{appendix1}
\input{T_terms}

\end{document}

%% file: table1.tex
\begin{table}
\begin{tabular}{|c|c|c|c|}
\hline
 $Q^m_l$ & Maximum $(\mu,t)$ & Minimum $(\mu,t)$  \\ \hline
$Q^0_0$ & 243.21 (1, 0.940)   &  1.62 (0.5, 1)  \\ \hline
$Q^0_2$ & 485.5 (1, 0.940)    &  0.58 (0.5, 1)    \\ \hline
$Q^1_2$ & 1.124 (0.974, 1)   & -68.32 (0.999, 0.956)  \\ \hline
$Q^2_2$ & 19.21 (0.999, 0.980)   & 0 (1., 0.5)     \\ \hline
$Q^0_4$ & 170.65 (1, 0.940)     & 0.045 (0.5, 1)     \\ \hline
$Q^1_4$ & 1.557 (0.812, 0.5)   & -40.84 (0.999, 0.942)     \\ \hline
$Q^2_4$ & 10.46 (0.999, 0.966)   &  -0.14 (0.938, 0.533)    \\ \hline
$Q^3_4$ & 0.226 (0.994, 1.0)   & -2.08 (0.999, 0.975)     \\ \hline
$Q^4_4$ & 1.22 (0.999, 1.)   & 0 (1., 0.50)     \\ \hline
$Q^0_6$ & 23.48 (1, 0.94)    & -0.83 (0.998, 0.980)    \\ \hline
$Q^1_6$ & 1.920$\times 10^{-4}$ (0.556, 1.)   & -7.06 (0.999, 0.952)     \\ \hline
$Q^2_6$ & 2.42 (0.999, 0.965)   & -0.18 (0.929, 0.603)     \\ \hline
$Q^3_6$ & 0.06 (0.972, 0.803)   & -0.43 (0.999, 0.979)     \\ \hline
$Q^4_6$ & 4.308$\times 10^{-3}$ (0.877, 1.)   & -0.09 (0.998, 0.977)     \\ \hline
$Q^5_6$ & 1.586$\times 10^{-3}$ (0.742, 1.)   & -9.172$\times 10^{-3}$ (0.971, 0.917)     \\ \hline
$Q^6_6$ & -4.27$\times 10^{-8}$ (0.997, 0.631)   & -7.27$\times 10^{-4}$ (0.724, 0.691)     \\ \hline
$Q^0_8$ & 1.15 (1, 0.940    ) &  -0.18 (0.998, 0.969)     \\ \hline
$Q^1_8$ & 8.49$\times 10^{-4}$ (0.626, 1.)   & -0.44 (0.999, 0.950)     \\ \hline
$Q^2_8$ & 0.20 (0.999, 0.967)   & -0.026 (0.920, 0.628)     \\ \hline
$Q^3_8$ & 0.021 (0.974, 0.867)   & -2.09$\times 10^{-4}$ (0.590, 1.)     \\ \hline
$Q^4_8$ & 0.005 (0.774, 0.646)   &-0.034 (0.999, 0.996)  \\ \hline
$Q^5_8$ & 0 (0.800, 0.625)   & -5.06$\times 10^{-3}$ (0.974, 0.964)    \\ \hline
$Q^6_8$ & -3.61$\times 10^{-9}$ (0.998, 0.502)   & -1.07$\times 10^{-3}$ (0.724, 0.691)     \\ \hline
\end{tabular}
\caption{It shows the maximum and minimum values of all the non-zero $Q^m_{\ell}$. }
\label{tab:1}
\end{table}

%% file: Sterms.tex
\begin{equation}
 A_0= 1+ 2 \bto/3  + \bto^2 /5
 \end{equation}
 \begin{equation}
 A_2=4(\bto/3  + \bto^2/7)
\end{equation}
\begin{equation}
 A_4=8 \bto^2/35
\end{equation}
\begin{equation}
[C_{12}]_{0} = [C_{23}]_{0} = [C_{31}]_{0} = -\frac{7 \bto ^2}{15} \,.
\label{eq:c1}
\end{equation}
\begin{equation}
[C_{12}]_{2}= [C_{31}]_{2} =-\frac{1}{6} \bto  (11 \bto +21)
\end{equation}
\begin{equation}
[C_{23}]_{2} =-\frac{1}{6} \bto  \left(11 \bto +6 (3 \bto +7) t^2-6 (3 \bto +7) \mu  t+21\right) \,.
\end{equation}
\begin{equation}
[C_{31}]_{4} =[C_{12}]_{4}  = -\frac{6 \bto ^2}{5}
\end{equation}
\begin{equation}
[C_{23}]_{4} =-\frac{1}{10} \bto ^2 \left(5 \left(7 \mu ^2+1\right) t^2-40 \mu  t+12\right)
\end{equation}
\begin{equation}
[D_{12}]_{2}^{1}  = \frac{1}{7} \sqrt{\frac{2}{3}} \bto  (3 \bto +7) \mu \,,
\end{equation}
\begin{dmath}
[D_{23}]_{2}^{1} =\sqrt{\frac{2}{3}} \frac{\bto }{7 s^2} \Big[\mu  \left(3 \bto +7 s^2\right)+(6 \bto +7) \mu  t^2-t \left(\bto  \left(6 \mu ^2+3\right)+7\right)\Big]
\end{dmath}
\begin{dmath}
[D_{31}]_{2}^{1} =\frac{\sqrt{\frac{2}{3}} \bto  (3 \bto +7) t (\mu  t-1)}{7 s^2} \,.
\end{dmath}
\begin{dmath}
[D_{12}]_{2}^{2} =\frac{\bto  (\bto +7)}{7 \sqrt{6}} \,,
\end{dmath}
\begin{dmath}
[D_{23}]_{2}^{2} =\frac{\bto  \left(\bto +7 s^2+(6 \bto +7) t^2-6 \bto  \mu  t\right)}{7 \sqrt{6} s^2} 
\end{dmath}
\begin{dmath}
[D_{31}]_{2}^{2}  = \frac{\bto  (\bto +7) t^2}{7 \sqrt{6} s^2} \,.
\end{dmath}
\begin{dmath}
[D_{12}]_{4}^{1}  = \frac{4 \bto ^2 \mu }{7 \sqrt{5}}
\end{dmath}
\begin{dmath}
[D_{23}]_{4}^{1}  = \frac{\bto ^2 (2 \mu  t-1) \left(\left(7 \mu ^2-3\right) t-4 \mu \right)}{7 \sqrt{5} s^2}
\end{dmath}
\begin{dmath}
[D_{31}]_{4}^{1}  = \frac{4 \bto ^2 t (\mu  t-1)}{7 \sqrt{5} s^2}
\end{dmath}
\begin{dmath}
[D_{12}]_{4}^{2}  = \frac{1}{7} \sqrt{\frac{2}{5}} \bto ^2
\end{dmath}
\begin{dmath}
[D_{23}]_{4}^{2}  = \frac{\sqrt{\frac{2}{5}} \bto ^2 \left(\left(7 \mu ^2-1\right) t^2-6 \mu  t+1\right)}{7 s^2}
\end{dmath}
\begin{dmath}
[D_{31}]_{4}^{2}  = \frac{\sqrt{\frac{2}{5}} \bto ^2 t^2}{7 s^2}
\end{dmath}
\begin{dmath}
[D_{23}]_{4}^{3}  = \frac{\bto ^2 t (2 \mu  t-1)}{\sqrt{35} s^2}
\end{dmath}
\begin{dmath}
[D_{23}]_{4}^{4}  = \frac{\bto ^2 t^2}{\sqrt{70} s^2}
\end{dmath}
$[D_{12}]_{4}^{3}, [D_{31}]_{4}^{3},D_{12}]_{4}^{4},[D_{31}]_{4}^{4}$ are  zero.

%% file: T_terms.tex
\begin{dmath}
\TB{0}{0}{12} =
-\frac{8}{315} \beta ^4 \mu ^4-\frac{8 \beta ^4 \mu ^2}{105}-\frac{\beta 
^4}{105}-\frac{12 \beta ^3 \mu ^2}{35}-\frac{3 \beta ^3}{35}-\frac{4 \beta ^2 
\mu ^2}{15}-\frac{\beta ^2}{3}-\frac{\beta }{3}+\frac{2}{63} \beta ^4 \mu ^3 
t+\frac{2 \beta ^4 \mu ^3}{63 t}+\frac{1}{42} \beta ^4 \mu  t+\frac{\beta ^4 \mu 
}{42 t}+\frac{2}{35} \beta ^3 \mu ^3 t+\frac{2 \beta ^3 \mu ^3}{35 
t}+\frac{11}{70} \beta ^3 \mu  t+\frac{11 \beta ^3 \mu }{70 t}+\frac{3}{10} 
\beta ^2 \mu  t+\frac{3 \beta ^2 \mu }{10 t}+\frac{\beta  \mu  t}{6}+\frac{\beta 
 \mu }{6 t}
\label{eq:apt1200}
\end{dmath}
\begin{dmath}
\TB{0}{0}{23} =
\frac{4 \beta ^4 \mu ^4}{105 s^4}+\frac{4 \beta ^4 \mu ^2}{35 s^4}+\frac{\beta 
^4}{70 s^4}+\frac{4 \beta ^3 \mu ^2}{35 s^4}+\frac{\beta ^3}{35 s^4}-\frac{\beta 
^2}{10 s^4}+\frac{\beta ^3 \mu  t^3}{14 s^4}+\frac{\beta ^2 \mu  t^3}{10 
s^4}+\frac{\beta ^4 \mu ^2 t^2}{21 s^4}+\frac{\beta ^4 t^2}{126 s^4}-\frac{4 
\beta ^3 \mu ^2 t^2}{35 s^4}-\frac{\beta ^3 t^2}{35 s^4}-\frac{\beta ^2 \mu ^2 
t^2}{5 s^4}-\frac{\beta ^2 t^2}{10 s^4}-\frac{2 \beta ^4 \mu ^3 t}{21 
s^4}-\frac{2 \beta ^4 \mu ^3}{63 s^4 t}-\frac{\beta ^4 \mu  t}{14 
s^4}-\frac{\beta ^4 \mu }{42 s^4 t}-\frac{\beta ^3 \mu }{14 s^4 t}+\frac{3 \beta 
^2 \mu  t}{10 s^4}+\frac{6 \beta ^3 \mu ^2}{35 s^2}+\frac{3 \beta ^3}{70 
s^2}+\frac{2 \beta ^2 \mu ^2}{15 s^2}+\frac{\beta ^2}{15 s^2}-\frac{\beta }{6 
s^2}-\frac{2 \beta ^3 \mu ^3}{35 s^2 t}-\frac{\beta ^3 \mu  t}{14 s^2}-\frac{3 
\beta ^3 \mu }{35 s^2 t}-\frac{\beta ^2 \mu }{5 s^2 t}+\frac{\beta  \mu  t}{6 
s^2}-\frac{\beta ^2 \mu }{10 t}-\frac{\beta  \mu }{6 t}
\label{eq:apt2300}
\end{dmath}
\begin{dmath}
\TB{0}{0}{31} =
-\frac{2 \beta ^4 \mu ^3 t^5}{63 s^4}-\frac{\beta ^4 \mu  t^5}{42 
s^4}-\frac{\beta ^3 \mu  t^5}{14 s^4}+\frac{4 \beta ^4 \mu ^4 t^4}{105 
s^4}+\frac{4 \beta ^4 \mu ^2 t^4}{35 s^4}+\frac{\beta ^4 t^4}{70 s^4}+\frac{4 
\beta ^3 \mu ^2 t^4}{35 s^4}+\frac{\beta ^3 t^4}{35 s^4}-\frac{\beta ^2 t^4}{10 
s^4}-\frac{2 \beta ^4 \mu ^3 t^3}{21 s^4}-\frac{\beta ^4 \mu  t^3}{14 
s^4}+\frac{3 \beta ^2 \mu  t^3}{10 s^4}+\frac{\beta ^4 \mu ^2 t^2}{21 
s^4}+\frac{\beta ^4 t^2}{126 s^4}-\frac{4 \beta ^3 \mu ^2 t^2}{35 
s^4}-\frac{\beta ^3 t^2}{35 s^4}-\frac{\beta ^2 \mu ^2 t^2}{5 s^4}-\frac{\beta 
^2 t^2}{10 s^4}+\frac{\beta ^3 \mu  t}{14 s^4}+\frac{\beta ^2 \mu  t}{10 
s^4}-\frac{2 \beta ^3 \mu ^3 t^3}{35 s^2}-\frac{3 \beta ^3 \mu  t^3}{35 
s^2}-\frac{\beta ^2 \mu  t^3}{5 s^2}+\frac{6 \beta ^3 \mu ^2 t^2}{35 
s^2}+\frac{3 \beta ^3 t^2}{70 s^2}+\frac{2 \beta ^2 \mu ^2 t^2}{15 
s^2}+\frac{\beta ^2 t^2}{15 s^2}-\frac{\beta  t^2}{6 s^2}-\frac{\beta ^3 \mu  
t}{14 s^2}+\frac{\beta  \mu  t}{6 s^2}-\frac{1}{10} \beta ^2 \mu  t-\frac{\beta  
\mu  t}{6}
\label{eq:apt3100}
\end{dmath}
\begin{dmath}
\TB{0}{2}{12} =
-\frac{136 \beta ^4 \mu ^4}{693}-\frac{46 \beta ^4 \mu ^2}{231}-\frac{2 \beta 
^4}{231}-\frac{2 \beta ^3 \mu ^4}{7}-\frac{15 \beta ^3 \mu ^2}{14}-\frac{\beta 
^3}{14}-\frac{31 \beta ^2 \mu ^2}{21}-\frac{5 \beta ^2}{21}-\frac{\beta  \mu 
^2}{2}-\frac{\beta }{6}+\frac{10}{231} \beta ^4 \mu ^5 t+\frac{95}{693} \beta ^4 
\mu ^3 t+\frac{205 \beta ^4 \mu ^3}{1386 t}+\frac{5}{231} \beta ^4 \mu  
t+\frac{25 \beta ^4 \mu }{462 t}+\frac{4}{7} \beta ^3 \mu ^3 t+\frac{\beta ^3 
\mu ^3}{3 t}+\frac{1}{7} \beta ^3 \mu  t+\frac{8 \beta ^3 \mu }{21 
t}+\frac{3}{7} \beta ^2 \mu ^3 t+\frac{3 \beta ^2 \mu ^3}{14 t}+\frac{3}{7} 
\beta ^2 \mu  t+\frac{9 \beta ^2 \mu }{14 t}+\frac{\beta  \mu  t}{3}+\frac{\beta 
 \mu }{3 t}
\label{eq:apt1220}
\end{dmath}
\begin{dmath}
\TB{0}{2}{23} =
\frac{68 \beta ^4 \mu ^4}{231 s^4}+\frac{23 \beta ^4 \mu ^2}{77 s^4}+\frac{\beta 
^4}{77 s^4}+\frac{3 \beta ^3 \mu ^2}{7 s^4}+\frac{\beta ^3}{21 s^4}-\frac{2 
\beta ^2}{7 s^4}+\frac{5 \beta ^3 \mu ^3 t^3}{21 s^4}+\frac{3 \beta ^2 \mu ^3 
t^3}{14 s^4}+\frac{\beta ^2 \mu  t^3}{14 s^4}+\frac{10 \beta ^4 \mu ^4 t^2}{77 
s^4}+\frac{5 \beta ^4 \mu ^2 t^2}{66 s^4}-\frac{5 \beta ^4 t^2}{1386 
s^4}-\frac{4 \beta ^3 \mu ^4 t^2}{21 s^4}-\frac{2 \beta ^3 \mu ^2 t^2}{7 
s^4}-\frac{11 \beta ^2 \mu ^2 t^2}{14 s^4}-\frac{\beta ^2 t^2}{14 s^4}-\frac{10 
\beta ^4 \mu ^5 t}{77 s^4}-\frac{95 \beta ^4 \mu ^3 t}{231 s^4}-\frac{205 \beta 
^4 \mu ^3}{1386 s^4 t}-\frac{5 \beta ^4 \mu  t}{77 s^4}-\frac{25 \beta ^4 \mu 
}{462 s^4 t}-\frac{5 \beta ^3 \mu }{21 s^4 t}+\frac{6 \beta ^2 \mu  t}{7 
s^4}+\frac{2 \beta ^3 \mu ^4}{7 s^2}+\frac{3 \beta ^3 \mu ^2}{7 s^2}+\frac{11 
\beta ^2 \mu ^2}{21 s^2}+\frac{\beta ^2}{21 s^2}-\frac{\beta }{3 s^2}-\frac{5 
\beta ^3 \mu ^3 t}{21 s^2}-\frac{\beta ^3 \mu ^3}{3 s^2 t}-\frac{\beta ^3 \mu 
}{7 s^2 t}-\frac{4 \beta ^2 \mu }{7 s^2 t}+\frac{\beta  \mu  t}{3 s^2}-\frac{3 
\beta ^2 \mu ^3}{14 t}-\frac{\beta ^2 \mu }{14 t}-\frac{\beta  \mu }{3 t}
\label{eq:apt2320}
\end{dmath}
\begin{dmath}
\TB{0}{2}{31} =
-\frac{10 \beta ^4 \mu ^5 t^5}{231 s^4}-\frac{95 \beta ^4 \mu ^3 t^5}{693 
s^4}-\frac{5 \beta ^4 \mu  t^5}{231 s^4}-\frac{5 \beta ^3 \mu ^3 t^5}{21 
s^4}+\frac{68 \beta ^4 \mu ^4 t^4}{231 s^4}+\frac{23 \beta ^4 \mu ^2 t^4}{77 
s^4}+\frac{\beta ^4 t^4}{77 s^4}+\frac{4 \beta ^3 \mu ^4 t^4}{21 s^4}+\frac{2 
\beta ^3 \mu ^2 t^4}{7 s^4}-\frac{3 \beta ^2 \mu ^2 t^4}{7 s^4}+\frac{\beta ^2 
t^4}{7 s^4}-\frac{205 \beta ^4 \mu ^3 t^3}{462 s^4}-\frac{25 \beta ^4 \mu  
t^3}{154 s^4}+\frac{9 \beta ^2 \mu ^3 t^3}{14 s^4}+\frac{3 \beta ^2 \mu  t^3}{14 
s^4}+\frac{85 \beta ^4 \mu ^2 t^2}{462 s^4}+\frac{25 \beta ^4 t^2}{1386 
s^4}-\frac{3 \beta ^3 \mu ^2 t^2}{7 s^4}-\frac{\beta ^3 t^2}{21 s^4}-\frac{11 
\beta ^2 \mu ^2 t^2}{14 s^4}-\frac{\beta ^2 t^2}{14 s^4}+\frac{5 \beta ^3 \mu  
t}{21 s^4}+\frac{2 \beta ^2 \mu  t}{7 s^4}-\frac{\beta ^3 \mu ^3 t^3}{3 
s^2}-\frac{\beta ^3 \mu  t^3}{7 s^2}-\frac{3 \beta ^2 \mu ^3 t^3}{7 
s^2}-\frac{\beta ^2 \mu  t^3}{7 s^2}+\frac{9 \beta ^3 \mu ^2 t^2}{14 
s^2}+\frac{\beta ^3 t^2}{14 s^2}+\frac{11 \beta ^2 \mu ^2 t^2}{21 
s^2}+\frac{\beta ^2 t^2}{21 s^2}-\frac{\beta  \mu ^2 t^2}{2 s^2}+\frac{\beta  
t^2}{6 s^2}-\frac{5 \beta ^3 \mu  t}{21 s^2}+\frac{\beta  \mu  t}{3 
s^2}-\frac{2}{7} \beta ^2 \mu  t-\frac{\beta  \mu  t}{3}
\label{eq:apt3120}
\end{dmath}
\begin{dmath}
\TB{1}{2}{12} =
\sqrt{1-\mu ^2} \Big\lbrace-\frac{40}{231} \sqrt{\frac{2}{3}} \beta ^4 \mu 
^3-\frac{10}{77} \sqrt{\frac{2}{3}} \beta ^4 \mu -\frac{2}{7} \sqrt{\frac{2}{3}} 
\beta ^3 \mu ^3-\frac{11 \beta ^3 \mu }{7 \sqrt{6}}-\frac{3}{7} \sqrt{6} \beta 
^2 \mu -\frac{\beta  \mu }{\sqrt{6}}+\frac{10}{231} \sqrt{\frac{2}{3}} \beta ^4 
\mu ^4 t+\frac{10}{77} \sqrt{\frac{2}{3}} \beta ^4 \mu ^2 t+\frac{5 
\sqrt{\frac{3}{2}} \beta ^4 \mu ^2}{77 t}+\frac{5 \beta ^4 t}{154 
\sqrt{6}}+\frac{5 \beta ^4}{154 \sqrt{6} t}+\frac{11}{21} \sqrt{\frac{2}{3}} 
\beta ^3 \mu ^2 t+\frac{2 \sqrt{\frac{2}{3}} \beta ^3 \mu ^2}{7 t}+\frac{11 
\beta ^3 t}{42 \sqrt{6}}+\frac{11 \beta ^3}{42 \sqrt{6} t}+\frac{1}{7} \sqrt{6} 
\beta ^2 \mu ^2 t+\frac{\sqrt{\frac{3}{2}} \beta ^2 \mu ^2}{7 t}+\frac{3}{14} 
\sqrt{\frac{3}{2}} \beta ^2 t+\frac{3 \sqrt{\frac{3}{2}} \beta ^2}{14 
t}+\frac{\beta  t}{2 \sqrt{6}}+\frac{\beta }{2 \sqrt{6} t} \Big\rbrace
\label{eq:apt1221}
\end{dmath}
\begin{dmath}
\TB{1}{2}{23} =
\sqrt{1-\mu ^2} \Big\lbrace\frac{20 \sqrt{\frac{2}{3}} \beta ^4 \mu ^3}{77 
s^4}+\frac{5 \sqrt{6} \beta ^4 \mu }{77 s^4}+\frac{5 \sqrt{\frac{2}{3}} \beta ^3 
\mu }{21 s^4}+\frac{5 \sqrt{\frac{2}{3}} \beta ^3 \mu ^2 t^3}{21 s^4}+\frac{5 
\beta ^3 t^3}{42 \sqrt{6} s^4}+\frac{\sqrt{\frac{3}{2}} \beta ^2 \mu ^2 t^3}{7 
s^4}+\frac{\sqrt{\frac{3}{2}} \beta ^2 t^3}{14 s^4}+\frac{10 \sqrt{\frac{2}{3}} 
\beta ^4 \mu ^3 t^2}{77 s^4}+\frac{5 \sqrt{\frac{3}{2}} \beta ^4 \mu  t^2}{77 
s^4}-\frac{4 \sqrt{\frac{2}{3}} \beta ^3 \mu ^3 t^2}{21 s^4}-\frac{2 
\sqrt{\frac{2}{3}} \beta ^3 \mu  t^2}{7 s^4}-\frac{3 \sqrt{\frac{3}{2}} \beta ^2 
\mu  t^2}{7 s^4}-\frac{10 \sqrt{\frac{2}{3}} \beta ^4 \mu ^4 t}{77 s^4}-\frac{10 
\sqrt{6} \beta ^4 \mu ^2 t}{77 s^4}-\frac{5 \sqrt{\frac{3}{2}} \beta ^4 \mu 
^2}{77 s^4 t}-\frac{5 \sqrt{\frac{3}{2}} \beta ^4 t}{154 s^4}-\frac{5 \beta 
^4}{154 \sqrt{6} s^4 t}-\frac{5 \beta ^3}{42 \sqrt{6} s^4 t}+\frac{3 
\sqrt{\frac{3}{2}} \beta ^2 t}{14 s^4}+\frac{2 \sqrt{\frac{2}{3}} \beta ^3 \mu 
^3}{7 s^2}+\frac{\sqrt{6} \beta ^3 \mu }{7 s^2}+\frac{\sqrt{6} \beta ^2 \mu }{7 
s^2}-\frac{5 \sqrt{\frac{2}{3}} \beta ^3 \mu ^2 t}{21 s^2}-\frac{2 
\sqrt{\frac{2}{3}} \beta ^3 \mu ^2}{7 s^2 t}-\frac{5 \beta ^3 t}{42 \sqrt{6} 
s^2}-\frac{\beta ^3}{7 \sqrt{6} s^2 t}-\frac{\sqrt{\frac{3}{2}} \beta ^2}{7 s^2 
t}+\frac{\beta  t}{2 \sqrt{6} s^2}-\frac{\sqrt{\frac{3}{2}} \beta ^2 \mu ^2}{7 
t}-\frac{\sqrt{\frac{3}{2}} \beta ^2}{14 t}-\frac{\beta }{2 \sqrt{6} t} 
\Big\rbrace
\label{eq:apt2321}
\end{dmath}
\begin{dmath}
\TB{1}{2}{31} =
\sqrt{1-\mu ^2} \Big\lbrace-\frac{10 \sqrt{\frac{2}{3}} \beta ^4 \mu ^4 t^5}{231 
s^4}-\frac{10 \sqrt{\frac{2}{3}} \beta ^4 \mu ^2 t^5}{77 s^4}-\frac{5 \beta ^4 
t^5}{154 \sqrt{6} s^4}-\frac{5 \sqrt{\frac{2}{3}} \beta ^3 \mu ^2 t^5}{21 
s^4}-\frac{5 \beta ^3 t^5}{42 \sqrt{6} s^4}+\frac{20 \sqrt{\frac{2}{3}} \beta ^4 
\mu ^3 t^4}{77 s^4}+\frac{5 \sqrt{6} \beta ^4 \mu  t^4}{77 s^4}+\frac{4 
\sqrt{\frac{2}{3}} \beta ^3 \mu ^3 t^4}{21 s^4}+\frac{2 \sqrt{\frac{2}{3}} \beta 
^3 \mu  t^4}{7 s^4}-\frac{\sqrt{6} \beta ^2 \mu  t^4}{7 s^4}-\frac{15 
\sqrt{\frac{3}{2}} \beta ^4 \mu ^2 t^3}{77 s^4}-\frac{5 \sqrt{\frac{3}{2}} \beta 
^4 t^3}{154 s^4}+\frac{3 \sqrt{\frac{3}{2}} \beta ^2 \mu ^2 t^3}{7 s^4}+\frac{3 
\sqrt{\frac{3}{2}} \beta ^2 t^3}{14 s^4}+\frac{5 \beta ^4 \mu  t^2}{33 \sqrt{6} 
s^4}-\frac{5 \sqrt{\frac{2}{3}} \beta ^3 \mu  t^2}{21 s^4}-\frac{3 
\sqrt{\frac{3}{2}} \beta ^2 \mu  t^2}{7 s^4}+\frac{5 \beta ^3 t}{42 \sqrt{6} 
s^4}+\frac{\sqrt{\frac{3}{2}} \beta ^2 t}{14 s^4}-\frac{2 \sqrt{\frac{2}{3}} 
\beta ^3 \mu ^2 t^3}{7 s^2}-\frac{\beta ^3 t^3}{7 \sqrt{6} s^2}-\frac{\sqrt{6} 
\beta ^2 \mu ^2 t^3}{7 s^2}-\frac{\sqrt{\frac{3}{2}} \beta ^2 t^3}{7 
s^2}+\frac{5 \beta ^3 \mu  t^2}{7 \sqrt{6} s^2}+\frac{\sqrt{6} \beta ^2 \mu  
t^2}{7 s^2}-\frac{\beta  \mu  t^2}{\sqrt{6} s^2}-\frac{5 \beta ^3 t}{42 \sqrt{6} 
s^2}+\frac{\beta  t}{2 \sqrt{6} s^2}-\frac{1}{14} \sqrt{\frac{3}{2}} \beta ^2 
t-\frac{\beta  t}{2 \sqrt{6}} \Big\rbrace
\label{eq:apt3121}
\end{dmath}
\begin{dmath}
\TB{2}{2}{12} =
\frac{4}{77} \sqrt{\frac{2}{3}} \beta ^4 \mu ^4-\frac{1}{77} \sqrt{6} \beta ^4 
\mu ^2-\frac{1}{77} \sqrt{\frac{2}{3}} \beta ^4+\frac{1}{7} \sqrt{\frac{2}{3}} 
\beta ^3 \mu ^4-\frac{\beta ^3 \mu ^2}{14 \sqrt{6}}-\frac{1}{14} 
\sqrt{\frac{3}{2}} \beta ^3+\frac{5 \beta ^2 \mu ^2}{7 \sqrt{6}}-\frac{5 \beta 
^2}{7 \sqrt{6}}+\frac{\beta  \mu ^2}{2 \sqrt{6}}-\frac{\beta }{2 
\sqrt{6}}-\frac{5}{231} \sqrt{\frac{2}{3}} \beta ^4 \mu ^5 t-\frac{5 \beta ^4 
\mu ^3 t}{231 \sqrt{6}}-\frac{5 \beta ^4 \mu ^3}{154 \sqrt{6} t}+\frac{5 \beta 
^4 \mu  t}{77 \sqrt{6}}+\frac{5 \beta ^4 \mu }{154 \sqrt{6} t}-\frac{4}{21} 
\sqrt{\frac{2}{3}} \beta ^3 \mu ^3 t-\frac{\beta ^3 \mu ^3}{7 \sqrt{6} 
t}+\frac{4}{21} \sqrt{\frac{2}{3}} \beta ^3 \mu  t+\frac{\beta ^3 \mu }{7 
\sqrt{6} t}-\frac{1}{7} \sqrt{\frac{3}{2}} \beta ^2 \mu ^3 
t-\frac{\sqrt{\frac{3}{2}} \beta ^2 \mu ^3}{14 t}+\frac{1}{7} \sqrt{\frac{3}{2}} 
\beta ^2 \mu  t+\frac{\sqrt{\frac{3}{2}} \beta ^2 \mu }{14 t}
\label{eq:apt1222}
\end{dmath}
\begin{dmath}
\TB{2}{2}{23} =
-\frac{2 \sqrt{6} \beta ^4 \mu ^4}{77 s^4}+\frac{3 \sqrt{\frac{3}{2}} \beta ^4 
\mu ^2}{77 s^4}+\frac{\sqrt{\frac{3}{2}} \beta ^4}{77 s^4}-\frac{\beta ^3 \mu 
^2}{21 \sqrt{6} s^4}+\frac{\beta ^3}{21 \sqrt{6} s^4}-\frac{5 \beta ^3 \mu ^3 
t^3}{21 \sqrt{6} s^4}+\frac{5 \beta ^3 \mu  t^3}{21 \sqrt{6} 
s^4}-\frac{\sqrt{\frac{3}{2}} \beta ^2 \mu ^3 t^3}{14 
s^4}+\frac{\sqrt{\frac{3}{2}} \beta ^2 \mu  t^3}{14 s^4}-\frac{5 
\sqrt{\frac{2}{3}} \beta ^4 \mu ^4 t^2}{77 s^4}+\frac{5 \sqrt{\frac{3}{2}} \beta 
^4 \mu ^2 t^2}{154 s^4}+\frac{5 \beta ^4 t^2}{154 \sqrt{6} s^4}+\frac{2 
\sqrt{\frac{2}{3}} \beta ^3 \mu ^4 t^2}{21 s^4}-\frac{\sqrt{\frac{2}{3}} \beta 
^3 \mu ^2 t^2}{21 s^4}-\frac{\sqrt{\frac{2}{3}} \beta ^3 t^2}{21 
s^4}+\frac{\sqrt{\frac{3}{2}} \beta ^2 \mu ^2 t^2}{14 
s^4}-\frac{\sqrt{\frac{3}{2}} \beta ^2 t^2}{14 s^4}+\frac{5 \sqrt{\frac{2}{3}} 
\beta ^4 \mu ^5 t}{77 s^4}+\frac{5 \beta ^4 \mu ^3 t}{77 \sqrt{6} s^4}+\frac{5 
\beta ^4 \mu ^3}{154 \sqrt{6} s^4 t}-\frac{5 \sqrt{\frac{3}{2}} \beta ^4 \mu  
t}{77 s^4}-\frac{5 \beta ^4 \mu }{154 \sqrt{6} s^4 t}-\frac{\sqrt{\frac{2}{3}} 
\beta ^3 \mu ^4}{7 s^2}+\frac{\beta ^3 \mu ^2}{7 \sqrt{6} s^2}+\frac{\beta ^3}{7 
\sqrt{6} s^2}-\frac{\beta ^2 \mu ^2}{7 \sqrt{6} s^2}+\frac{\beta ^2}{7 \sqrt{6} 
s^2}+\frac{5 \beta ^3 \mu ^3 t}{21 \sqrt{6} s^2}+\frac{\beta ^3 \mu ^3}{7 
\sqrt{6} s^2 t}-\frac{5 \beta ^3 \mu  t}{21 \sqrt{6} s^2}-\frac{\beta ^3 \mu }{7 
\sqrt{6} s^2 t}+\frac{\sqrt{\frac{3}{2}} \beta ^2 \mu ^3}{14 
t}-\frac{\sqrt{\frac{3}{2}} \beta ^2 \mu }{14 t}
\label{eq:apt2322}
\end{dmath}
\begin{dmath}
\TB{2}{2}{31} =
\frac{5 \sqrt{\frac{2}{3}} \beta ^4 \mu ^5 t^5}{231 s^4}+\frac{5 \beta ^4 \mu ^3 
t^5}{231 \sqrt{6} s^4}-\frac{5 \beta ^4 \mu  t^5}{77 \sqrt{6} s^4}+\frac{5 \beta 
^3 \mu ^3 t^5}{21 \sqrt{6} s^4}-\frac{5 \beta ^3 \mu  t^5}{21 \sqrt{6} 
s^4}-\frac{2 \sqrt{6} \beta ^4 \mu ^4 t^4}{77 s^4}+\frac{3 \sqrt{\frac{3}{2}} 
\beta ^4 \mu ^2 t^4}{77 s^4}+\frac{\sqrt{\frac{3}{2}} \beta ^4 t^4}{77 
s^4}-\frac{2 \sqrt{\frac{2}{3}} \beta ^3 \mu ^4 t^4}{21 
s^4}+\frac{\sqrt{\frac{2}{3}} \beta ^3 \mu ^2 t^4}{21 
s^4}+\frac{\sqrt{\frac{2}{3}} \beta ^3 t^4}{21 s^4}+\frac{\sqrt{\frac{3}{2}} 
\beta ^2 \mu ^2 t^4}{7 s^4}-\frac{\sqrt{\frac{3}{2}} \beta ^2 t^4}{7 
s^4}+\frac{5 \sqrt{\frac{3}{2}} \beta ^4 \mu ^3 t^3}{154 s^4}-\frac{5 
\sqrt{\frac{3}{2}} \beta ^4 \mu  t^3}{154 s^4}-\frac{3 \sqrt{\frac{3}{2}} \beta 
^2 \mu ^3 t^3}{14 s^4}+\frac{3 \sqrt{\frac{3}{2}} \beta ^2 \mu  t^3}{14 
s^4}-\frac{5 \beta ^4 \mu ^2 t^2}{462 \sqrt{6} s^4}+\frac{5 \beta ^4 t^2}{462 
\sqrt{6} s^4}+\frac{\beta ^3 \mu ^2 t^2}{21 \sqrt{6} s^4}-\frac{\beta ^3 t^2}{21 
\sqrt{6} s^4}+\frac{\sqrt{\frac{3}{2}} \beta ^2 \mu ^2 t^2}{14 
s^4}-\frac{\sqrt{\frac{3}{2}} \beta ^2 t^2}{14 s^4}+\frac{\beta ^3 \mu ^3 t^3}{7 
\sqrt{6} s^2}-\frac{\beta ^3 \mu  t^3}{7 \sqrt{6} s^2}+\frac{\sqrt{\frac{3}{2}} 
\beta ^2 \mu ^3 t^3}{7 s^2}-\frac{\sqrt{\frac{3}{2}} \beta ^2 \mu  t^3}{7 
s^2}-\frac{\beta ^3 \mu ^2 t^2}{14 \sqrt{6} s^2}+\frac{\beta ^3 t^2}{14 \sqrt{6} 
s^2}-\frac{\beta ^2 \mu ^2 t^2}{7 \sqrt{6} s^2}+\frac{\beta ^2 t^2}{7 \sqrt{6} 
s^2}+\frac{\beta  \mu ^2 t^2}{2 \sqrt{6} s^2}-\frac{\beta  t^2}{2 \sqrt{6} s^2}
\label{eq:apt3122}
\end{dmath}
\begin{dmath}
\TB{0}{4}{12} =
-\frac{1929 \beta ^4 \mu ^4}{5005}+\frac{138 \beta ^4 \mu ^2}{5005}+\frac{111 
\beta ^4}{5005}-\frac{117 \beta ^3 \mu ^4}{154}-\frac{123 \beta ^3 \mu 
^2}{385}+\frac{111 \beta ^3}{770}-\frac{\beta ^2 \mu ^4}{2}-\frac{9 \beta ^2 \mu 
^2}{35}+\frac{\beta ^2}{14}+\frac{345 \beta ^4 \mu ^5 t}{2002}+\frac{51 \beta ^4 
\mu ^3 t}{1001}+\frac{186 \beta ^4 \mu ^3}{1001 t}-\frac{111 \beta ^4 \mu  
t}{2002}-\frac{18 \beta ^4 \mu }{1001 t}+\frac{5}{22} \beta ^3 \mu ^5 
t+\frac{193}{385} \beta ^3 \mu ^3 t+\frac{24 \beta ^3 \mu ^3}{55 
t}-\frac{201}{770} \beta ^3 \mu  t+\frac{12 \beta ^3 \mu }{385 t}+\frac{4}{7} 
\beta ^2 \mu ^3 t+\frac{2 \beta ^2 \mu ^3}{7 t}-\frac{8}{35} \beta ^2 \mu  
t+\frac{2 \beta ^2 \mu }{35 t}
\label{eq:apt1240}
\end{dmath}
\begin{dmath}
\TB{0}{4}{23} =
\frac{5787 \beta ^4 \mu ^4}{10010 s^4}-\frac{207 \beta ^4 \mu ^2}{5005 
s^4}-\frac{333 \beta ^4}{10010 s^4}+\frac{136 \beta ^3 \mu ^2}{385 s^4}-\frac{16 
\beta ^3}{385 s^4}-\frac{4 \beta ^2}{35 s^4}+\frac{5 \beta ^3 \mu ^5 t^3}{22 
s^4}+\frac{5 \beta ^3 \mu ^3 t^3}{77 s^4}-\frac{3 \beta ^3 \mu  t^3}{22 
s^4}+\frac{2 \beta ^2 \mu ^3 t^3}{7 s^4}-\frac{6 \beta ^2 \mu  t^3}{35 
s^4}+\frac{15 \beta ^4 \mu ^6 t^2}{143 s^4}+\frac{405 \beta ^4 \mu ^4 t^2}{2002 
s^4}-\frac{18 \beta ^4 \mu ^2 t^2}{143 s^4}-\frac{27 \beta ^4 t^2}{2002 
s^4}-\frac{39 \beta ^3 \mu ^4 t^2}{77 s^4}+\frac{54 \beta ^3 \mu ^2 t^2}{385 
s^4}+\frac{3 \beta ^3 t^2}{55 s^4}-\frac{18 \beta ^2 \mu ^2 t^2}{35 s^4}+\frac{6 
\beta ^2 t^2}{35 s^4}-\frac{1035 \beta ^4 \mu ^5 t}{2002 s^4}-\frac{153 \beta ^4 
\mu ^3 t}{1001 s^4}-\frac{186 \beta ^4 \mu ^3}{1001 s^4 t}+\frac{333 \beta ^4 
\mu  t}{2002 s^4}+\frac{18 \beta ^4 \mu }{1001 s^4 t}-\frac{12 \beta ^3 \mu }{77 
s^4 t}+\frac{12 \beta ^2 \mu  t}{35 s^4}+\frac{117 \beta ^3 \mu ^4}{154 
s^2}-\frac{81 \beta ^3 \mu ^2}{385 s^2}-\frac{9 \beta ^3}{110 s^2}+\frac{12 
\beta ^2 \mu ^2}{35 s^2}-\frac{4 \beta ^2}{35 s^2}-\frac{5 \beta ^3 \mu ^5 t}{22 
s^2}-\frac{5 \beta ^3 \mu ^3 t}{77 s^2}-\frac{24 \beta ^3 \mu ^3}{55 s^2 
t}+\frac{3 \beta ^3 \mu  t}{22 s^2}+\frac{48 \beta ^3 \mu }{385 s^2 t}-\frac{8 
\beta ^2 \mu }{35 s^2 t}-\frac{2 \beta ^2 \mu ^3}{7 t}+\frac{6 \beta ^2 \mu }{35 
t}
\label{eq:apt2340}
\end{dmath}
\begin{dmath}
\TB{0}{4}{31} =
-\frac{345 \beta ^4 \mu ^5 t^5}{2002 s^4}-\frac{51 \beta ^4 \mu ^3 t^5}{1001 
s^4}+\frac{111 \beta ^4 \mu  t^5}{2002 s^4}-\frac{5 \beta ^3 \mu ^5 t^5}{22 
s^4}-\frac{5 \beta ^3 \mu ^3 t^5}{77 s^4}+\frac{3 \beta ^3 \mu  t^5}{22 
s^4}+\frac{5787 \beta ^4 \mu ^4 t^4}{10010 s^4}-\frac{207 \beta ^4 \mu ^2 
t^4}{5005 s^4}-\frac{333 \beta ^4 t^4}{10010 s^4}+\frac{39 \beta ^3 \mu ^4 
t^4}{77 s^4}-\frac{54 \beta ^3 \mu ^2 t^4}{385 s^4}-\frac{3 \beta ^3 t^4}{55 
s^4}-\frac{\beta ^2 \mu ^4 t^4}{2 s^4}+\frac{3 \beta ^2 \mu ^2 t^4}{7 
s^4}-\frac{3 \beta ^2 t^4}{70 s^4}-\frac{558 \beta ^4 \mu ^3 t^3}{1001 
s^4}+\frac{54 \beta ^4 \mu  t^3}{1001 s^4}+\frac{6 \beta ^2 \mu ^3 t^3}{7 
s^4}-\frac{18 \beta ^2 \mu  t^3}{35 s^4}+\frac{174 \beta ^4 \mu ^2 t^2}{1001 
s^4}-\frac{6 \beta ^4 t^2}{1001 s^4}-\frac{136 \beta ^3 \mu ^2 t^2}{385 
s^4}+\frac{16 \beta ^3 t^2}{385 s^4}-\frac{18 \beta ^2 \mu ^2 t^2}{35 
s^4}+\frac{6 \beta ^2 t^2}{35 s^4}+\frac{12 \beta ^3 \mu  t}{77 s^4}+\frac{4 
\beta ^2 \mu  t}{35 s^4}-\frac{24 \beta ^3 \mu ^3 t^3}{55 s^2}+\frac{48 \beta ^3 
\mu  t^3}{385 s^2}-\frac{4 \beta ^2 \mu ^3 t^3}{7 s^2}+\frac{12 \beta ^2 \mu  
t^3}{35 s^2}+\frac{204 \beta ^3 \mu ^2 t^2}{385 s^2}-\frac{24 \beta ^3 t^2}{385 
s^2}+\frac{12 \beta ^2 \mu ^2 t^2}{35 s^2}-\frac{4 \beta ^2 t^2}{35 
s^2}-\frac{12 \beta ^3 \mu  t}{77 s^2}-\frac{4}{35} \beta ^2 \mu  t
\label{eq:apt3140}
\end{dmath}
\begin{dmath}
\TB{1}{4}{12} =
\sqrt{1-\mu ^2} \Big\lbrace-\frac{138 \sqrt{5} \beta ^4 \mu ^3}{1001}-\frac{30 
\sqrt{5} \beta ^4 \mu }{1001}-\frac{111 \beta ^3 \mu ^3}{77 \sqrt{5}}-\frac{69 
\beta ^3 \mu }{77 \sqrt{5}}-\frac{\beta ^2 \mu ^3}{\sqrt{5}}-\frac{1}{7} 
\sqrt{5} \beta ^2 \mu +\frac{6}{91} \sqrt{5} \beta ^4 \mu ^4 t+\frac{81 \sqrt{5} 
\beta ^4 \mu ^2 t}{2002}+\frac{237 \sqrt{5} \beta ^4 \mu ^2}{4004 t}-\frac{3 
\sqrt{5} \beta ^4 t}{2002}+\frac{15 \sqrt{5} \beta ^4}{4004 t}+\frac{1}{11} 
\sqrt{5} \beta ^3 \mu ^4 t+\frac{81 \beta ^3 \mu ^2 t}{77 \sqrt{5}}+\frac{117 
\beta ^3 \mu ^2}{154 \sqrt{5} t}-\frac{6 \beta ^3 t}{77 \sqrt{5}}+\frac{23 \beta 
^3}{154 \sqrt{5} t}+\frac{3}{14} \sqrt{5} \beta ^2 \mu ^2 t+\frac{3 \sqrt{5} 
\beta ^2 \mu ^2}{28 t}-\frac{\beta ^2 t}{14 \sqrt{5}}+\frac{\sqrt{5} \beta 
^2}{28 t} \Big\rbrace
\label{eq:apt1241}
\end{dmath}
\begin{dmath}
\TB{1}{4}{23} =
\sqrt{1-\mu ^2} \Big\lbrace\frac{207 \sqrt{5} \beta ^4 \mu ^3}{1001 
s^4}+\frac{45 \sqrt{5} \beta ^4 \mu }{1001 s^4}+\frac{8 \sqrt{5} \beta ^3 \mu 
}{77 s^4}+\frac{\sqrt{5} \beta ^3 \mu ^4 t^3}{11 s^4}+\frac{9 \sqrt{5} \beta ^3 
\mu ^2 t^3}{154 s^4}-\frac{3 \sqrt{5} \beta ^3 t^3}{154 s^4}+\frac{3 \sqrt{5} 
\beta ^2 \mu ^2 t^3}{28 s^4}-\frac{3 \beta ^2 t^3}{28 \sqrt{5} s^4}+\frac{6 
\sqrt{5} \beta ^4 \mu ^5 t^2}{143 s^4}+\frac{93 \sqrt{5} \beta ^4 \mu ^3 
t^2}{1001 s^4}-\frac{9 \sqrt{5} \beta ^4 \mu  t^2}{1001 s^4}-\frac{74 \beta ^3 
\mu ^3 t^2}{77 \sqrt{5} s^4}-\frac{6 \beta ^3 \mu  t^2}{77 \sqrt{5} s^4}-\frac{6 
\beta ^2 \mu  t^2}{7 \sqrt{5} s^4}-\frac{18 \sqrt{5} \beta ^4 \mu ^4 t}{91 
s^4}-\frac{243 \sqrt{5} \beta ^4 \mu ^2 t}{2002 s^4}-\frac{237 \sqrt{5} \beta ^4 
\mu ^2}{4004 s^4 t}+\frac{9 \sqrt{5} \beta ^4 t}{2002 s^4}-\frac{15 \sqrt{5} 
\beta ^4}{4004 s^4 t}-\frac{2 \sqrt{5} \beta ^3}{77 s^4 t}+\frac{3 \beta ^2 t}{7 
\sqrt{5} s^4}+\frac{111 \beta ^3 \mu ^3}{77 \sqrt{5} s^2}+\frac{9 \beta ^3 \mu 
}{77 \sqrt{5} s^2}+\frac{4 \beta ^2 \mu }{7 \sqrt{5} s^2}-\frac{\sqrt{5} \beta 
^3 \mu ^4 t}{11 s^2}-\frac{9 \sqrt{5} \beta ^3 \mu ^2 t}{154 s^2}-\frac{117 
\beta ^3 \mu ^2}{154 \sqrt{5} s^2 t}+\frac{3 \sqrt{5} \beta ^3 t}{154 
s^2}-\frac{3 \beta ^3}{154 \sqrt{5} s^2 t}-\frac{2 \beta ^2}{7 \sqrt{5} s^2 
t}-\frac{3 \sqrt{5} \beta ^2 \mu ^2}{28 t}+\frac{3 \beta ^2}{28 \sqrt{5} t} 
\Big\rbrace
\label{eq:apt2341}
\end{dmath}
\begin{dmath}
\TB{1}{4}{31} =
\sqrt{1-\mu ^2} \Big\lbrace-\frac{6 \sqrt{5} \beta ^4 \mu ^4 t^5}{91 
s^4}-\frac{81 \sqrt{5} \beta ^4 \mu ^2 t^5}{2002 s^4}+\frac{3 \sqrt{5} \beta ^4 
t^5}{2002 s^4}-\frac{\sqrt{5} \beta ^3 \mu ^4 t^5}{11 s^4}-\frac{9 \sqrt{5} 
\beta ^3 \mu ^2 t^5}{154 s^4}+\frac{3 \sqrt{5} \beta ^3 t^5}{154 s^4}+\frac{207 
\sqrt{5} \beta ^4 \mu ^3 t^4}{1001 s^4}+\frac{45 \sqrt{5} \beta ^4 \mu  
t^4}{1001 s^4}+\frac{74 \beta ^3 \mu ^3 t^4}{77 \sqrt{5} s^4}+\frac{6 \beta ^3 
\mu  t^4}{77 \sqrt{5} s^4}-\frac{\beta ^2 \mu ^3 t^4}{\sqrt{5} s^4}+\frac{3 
\beta ^2 \mu  t^4}{7 \sqrt{5} s^4}-\frac{711 \sqrt{5} \beta ^4 \mu ^2 t^3}{4004 
s^4}-\frac{45 \sqrt{5} \beta ^4 t^3}{4004 s^4}+\frac{9 \sqrt{5} \beta ^2 \mu ^2 
t^3}{28 s^4}-\frac{9 \beta ^2 t^3}{28 \sqrt{5} s^4}+\frac{6 \sqrt{5} \beta ^4 
\mu  t^2}{143 s^4}-\frac{8 \sqrt{5} \beta ^3 \mu  t^2}{77 s^4}-\frac{6 \beta ^2 
\mu  t^2}{7 \sqrt{5} s^4}+\frac{2 \sqrt{5} \beta ^3 t}{77 s^4}+\frac{\beta ^2 
t}{7 \sqrt{5} s^4}-\frac{117 \beta ^3 \mu ^2 t^3}{154 \sqrt{5} s^2}-\frac{3 
\beta ^3 t^3}{154 \sqrt{5} s^2}-\frac{3 \sqrt{5} \beta ^2 \mu ^2 t^3}{14 
s^2}+\frac{3 \beta ^2 t^3}{14 \sqrt{5} s^2}+\frac{12 \sqrt{5} \beta ^3 \mu  
t^2}{77 s^2}+\frac{4 \beta ^2 \mu  t^2}{7 \sqrt{5} s^2}-\frac{2 \sqrt{5} \beta 
^3 t}{77 s^2}-\frac{\beta ^2 t}{7 \sqrt{5}} \Big\rbrace
\label{eq:apt3141}
\end{dmath}
\begin{dmath}
\TB{2}{4}{12} =
\frac{237 \sqrt{\frac{2}{5}} \beta ^4 \mu ^4}{1001}-\frac{204 \sqrt{\frac{2}{5}} 
\beta ^4 \mu ^2}{1001}-\frac{3}{91} \sqrt{\frac{2}{5}} \beta ^4+\frac{93 \beta 
^3 \mu ^4}{77 \sqrt{10}}-\frac{6}{77} \sqrt{10} \beta ^3 \mu ^2-\frac{3 \beta 
^3}{7 \sqrt{10}}+\frac{\beta ^2 \mu ^4}{\sqrt{10}}-\frac{2}{7} 
\sqrt{\frac{2}{5}} \beta ^2 \mu ^2-\frac{3 \beta ^2}{7 \sqrt{10}}-\frac{57 
\sqrt{\frac{5}{2}} \beta ^4 \mu ^5 t}{1001}+\frac{12 \sqrt{10} \beta ^4 \mu ^3 
t}{1001}-\frac{27 \sqrt{\frac{5}{2}} \beta ^4 \mu ^3}{1001 t}+\frac{3}{91} 
\sqrt{\frac{5}{2}} \beta ^4 \mu  t+\frac{27 \sqrt{\frac{5}{2}} \beta ^4 \mu 
}{1001 t}-\frac{1}{11} \sqrt{\frac{5}{2}} \beta ^3 \mu ^5 t-\frac{13}{77} 
\sqrt{\frac{2}{5}} \beta ^3 \mu ^3 t-\frac{18 \sqrt{\frac{2}{5}} \beta ^3 \mu 
^3}{77 t}+\frac{61 \beta ^3 \mu  t}{77 \sqrt{10}}+\frac{18 \sqrt{\frac{2}{5}} 
\beta ^3 \mu }{77 t}-\frac{3}{7} \sqrt{\frac{2}{5}} \beta ^2 \mu ^3 t-\frac{3 
\beta ^2 \mu ^3}{7 \sqrt{10} t}+\frac{3}{7} \sqrt{\frac{2}{5}} \beta ^2 \mu  
t+\frac{3 \beta ^2 \mu }{7 \sqrt{10} t}
\label{eq:apt1242}
\end{dmath}
\begin{dmath}
\TB{2}{4}{23} =
-\frac{711 \beta ^4 \mu ^4}{1001 \sqrt{10} s^4}+\frac{306 \sqrt{\frac{2}{5}} 
\beta ^4 \mu ^2}{1001 s^4}+\frac{9 \beta ^4}{91 \sqrt{10} s^4}-\frac{6 
\sqrt{\frac{2}{5}} \beta ^3 \mu ^2}{77 s^4}+\frac{6 \sqrt{\frac{2}{5}} \beta 
^3}{77 s^4}-\frac{\sqrt{\frac{5}{2}} \beta ^3 \mu ^5 t^3}{11 
s^4}+\frac{\sqrt{10} \beta ^3 \mu ^3 t^3}{77 s^4}+\frac{5 \sqrt{\frac{5}{2}} 
\beta ^3 \mu  t^3}{77 s^4}-\frac{3 \beta ^2 \mu ^3 t^3}{7 \sqrt{10} s^4}+\frac{3 
\beta ^2 \mu  t^3}{7 \sqrt{10} s^4}-\frac{3 \sqrt{10} \beta ^4 \mu ^6 t^2}{143 
s^4}-\frac{45 \sqrt{\frac{5}{2}} \beta ^4 \mu ^4 t^2}{1001 s^4}+\frac{81 
\sqrt{\frac{5}{2}} \beta ^4 \mu ^2 t^2}{1001 s^4}+\frac{3 \sqrt{10} \beta ^4 
t^2}{1001 s^4}+\frac{31 \sqrt{\frac{2}{5}} \beta ^3 \mu ^4 t^2}{77 s^4}-\frac{26 
\sqrt{\frac{2}{5}} \beta ^3 \mu ^2 t^2}{77 s^4}-\frac{\sqrt{10} \beta ^3 t^2}{77 
s^4}+\frac{3 \beta ^2 \mu ^2 t^2}{7 \sqrt{10} s^4}-\frac{3 \beta ^2 t^2}{7 
\sqrt{10} s^4}+\frac{171 \sqrt{\frac{5}{2}} \beta ^4 \mu ^5 t}{1001 
s^4}-\frac{36 \sqrt{10} \beta ^4 \mu ^3 t}{1001 s^4}+\frac{27 \sqrt{\frac{5}{2}} 
\beta ^4 \mu ^3}{1001 s^4 t}-\frac{9 \sqrt{\frac{5}{2}} \beta ^4 \mu  t}{91 
s^4}-\frac{27 \sqrt{\frac{5}{2}} \beta ^4 \mu }{1001 s^4 t}-\frac{93 \beta ^3 
\mu ^4}{77 \sqrt{10} s^2}+\frac{39 \sqrt{\frac{2}{5}} \beta ^3 \mu ^2}{77 
s^2}+\frac{3 \sqrt{\frac{5}{2}} \beta ^3}{77 s^2}-\frac{\sqrt{\frac{2}{5}} \beta 
^2 \mu ^2}{7 s^2}+\frac{\sqrt{\frac{2}{5}} \beta ^2}{7 
s^2}+\frac{\sqrt{\frac{5}{2}} \beta ^3 \mu ^5 t}{11 s^2}-\frac{\sqrt{10} \beta 
^3 \mu ^3 t}{77 s^2}+\frac{18 \sqrt{\frac{2}{5}} \beta ^3 \mu ^3}{77 s^2 
t}-\frac{5 \sqrt{\frac{5}{2}} \beta ^3 \mu  t}{77 s^2}-\frac{18 
\sqrt{\frac{2}{5}} \beta ^3 \mu }{77 s^2 t}+\frac{3 \beta ^2 \mu ^3}{7 \sqrt{10} 
t}-\frac{3 \beta ^2 \mu }{7 \sqrt{10} t}
\label{eq:apt2342}
\end{dmath}
\begin{dmath}
\TB{2}{4}{31} =
\frac{57 \sqrt{\frac{5}{2}} \beta ^4 \mu ^5 t^5}{1001 s^4}-\frac{12 \sqrt{10} 
\beta ^4 \mu ^3 t^5}{1001 s^4}-\frac{3 \sqrt{\frac{5}{2}} \beta ^4 \mu  t^5}{91 
s^4}+\frac{\sqrt{\frac{5}{2}} \beta ^3 \mu ^5 t^5}{11 s^4}-\frac{\sqrt{10} \beta 
^3 \mu ^3 t^5}{77 s^4}-\frac{5 \sqrt{\frac{5}{2}} \beta ^3 \mu  t^5}{77 
s^4}-\frac{711 \beta ^4 \mu ^4 t^4}{1001 \sqrt{10} s^4}+\frac{306 
\sqrt{\frac{2}{5}} \beta ^4 \mu ^2 t^4}{1001 s^4}+\frac{9 \beta ^4 t^4}{91 
\sqrt{10} s^4}-\frac{31 \sqrt{\frac{2}{5}} \beta ^3 \mu ^4 t^4}{77 s^4}+\frac{26 
\sqrt{\frac{2}{5}} \beta ^3 \mu ^2 t^4}{77 s^4}+\frac{\sqrt{10} \beta ^3 t^4}{77 
s^4}+\frac{\beta ^2 \mu ^4 t^4}{\sqrt{10} s^4}-\frac{4 \sqrt{\frac{2}{5}} \beta 
^2 \mu ^2 t^4}{7 s^4}+\frac{\beta ^2 t^4}{7 \sqrt{10} s^4}+\frac{81 
\sqrt{\frac{5}{2}} \beta ^4 \mu ^3 t^3}{1001 s^4}-\frac{81 \sqrt{\frac{5}{2}} 
\beta ^4 \mu  t^3}{1001 s^4}-\frac{9 \beta ^2 \mu ^3 t^3}{7 \sqrt{10} 
s^4}+\frac{9 \beta ^2 \mu  t^3}{7 \sqrt{10} s^4}-\frac{9 \sqrt{\frac{5}{2}} 
\beta ^4 \mu ^2 t^2}{1001 s^4}+\frac{9 \sqrt{\frac{5}{2}} \beta ^4 t^2}{1001 
s^4}+\frac{6 \sqrt{\frac{2}{5}} \beta ^3 \mu ^2 t^2}{77 s^4}-\frac{6 
\sqrt{\frac{2}{5}} \beta ^3 t^2}{77 s^4}+\frac{3 \beta ^2 \mu ^2 t^2}{7 
\sqrt{10} s^4}-\frac{3 \beta ^2 t^2}{7 \sqrt{10} s^4}+\frac{18 
\sqrt{\frac{2}{5}} \beta ^3 \mu ^3 t^3}{77 s^2}-\frac{18 \sqrt{\frac{2}{5}} 
\beta ^3 \mu  t^3}{77 s^2}+\frac{3 \sqrt{\frac{2}{5}} \beta ^2 \mu ^3 t^3}{7 
s^2}-\frac{3 \sqrt{\frac{2}{5}} \beta ^2 \mu  t^3}{7 s^2}-\frac{9 
\sqrt{\frac{2}{5}} \beta ^3 \mu ^2 t^2}{77 s^2}+\frac{9 \sqrt{\frac{2}{5}} \beta 
^3 t^2}{77 s^2}-\frac{\sqrt{\frac{2}{5}} \beta ^2 \mu ^2 t^2}{7 
s^2}+\frac{\sqrt{\frac{2}{5}} \beta ^2 t^2}{7 s^2}
\label{eq:apt3142}
\end{dmath}
\begin{dmath}
\TB{3}{4}{12} =
\sqrt{1-\mu ^2} \Big\lbrace\frac{6}{143} \sqrt{\frac{5}{7}} \beta ^4 \mu 
^3-\frac{6}{143} \sqrt{\frac{5}{7}} \beta ^4 \mu +\frac{9 \beta ^3 \mu ^3}{11 
\sqrt{35}}-\frac{9 \beta ^3 \mu }{11 \sqrt{35}}+\frac{\beta ^2 \mu 
^3}{\sqrt{35}}-\frac{\beta ^2 \mu }{\sqrt{35}}-\frac{6}{143} \sqrt{\frac{5}{7}} 
\beta ^4 \mu ^4 t+\frac{9}{286} \sqrt{\frac{5}{7}} \beta ^4 \mu ^2 t-\frac{3 
\sqrt{\frac{5}{7}} \beta ^4 \mu ^2}{572 t}+\frac{3}{286} \sqrt{\frac{5}{7}} 
\beta ^4 t+\frac{3 \sqrt{\frac{5}{7}} \beta ^4}{572 t}-\frac{1}{11} 
\sqrt{\frac{5}{7}} \beta ^3 \mu ^4 t+\frac{\beta ^3 \mu ^2 t}{11 
\sqrt{35}}-\frac{3 \beta ^3 \mu ^2}{22 \sqrt{35} t}+\frac{4 \beta ^3 t}{11 
\sqrt{35}}+\frac{3 \beta ^3}{22 \sqrt{35} t}-\frac{\beta ^2 \mu ^2 t}{2 
\sqrt{35}}-\frac{\beta ^2 \mu ^2}{4 \sqrt{35} t}+\frac{\beta ^2 t}{2 
\sqrt{35}}+\frac{\beta ^2}{4 \sqrt{35} t} \Big\rbrace
\label{eq:apt1243}
\end{dmath}
\begin{dmath}
\TB{3}{4}{23} =
\sqrt{1-\mu ^2} \Big\lbrace-\frac{9 \sqrt{\frac{5}{7}} \beta ^4 \mu ^3}{143 
s^4}+\frac{9 \sqrt{\frac{5}{7}} \beta ^4 \mu }{143 s^4}-\frac{\sqrt{\frac{5}{7}} 
\beta ^3 \mu ^4 t^3}{11 s^4}+\frac{\sqrt{\frac{5}{7}} \beta ^3 \mu ^2 t^3}{22 
s^4}+\frac{\sqrt{\frac{5}{7}} \beta ^3 t^3}{22 s^4}-\frac{\beta ^2 \mu ^2 t^3}{4 
\sqrt{35} s^4}+\frac{\beta ^2 t^3}{4 \sqrt{35} s^4}-\frac{6 \sqrt{\frac{5}{7}} 
\beta ^4 \mu ^5 t^2}{143 s^4}-\frac{3 \sqrt{\frac{5}{7}} \beta ^4 \mu ^3 
t^2}{143 s^4}+\frac{9 \sqrt{\frac{5}{7}} \beta ^4 \mu  t^2}{143 s^4}+\frac{6 
\beta ^3 \mu ^3 t^2}{11 \sqrt{35} s^4}-\frac{6 \beta ^3 \mu  t^2}{11 \sqrt{35} 
s^4}+\frac{18 \sqrt{\frac{5}{7}} \beta ^4 \mu ^4 t}{143 s^4}-\frac{27 
\sqrt{\frac{5}{7}} \beta ^4 \mu ^2 t}{286 s^4}+\frac{3 \sqrt{\frac{5}{7}} \beta 
^4 \mu ^2}{572 s^4 t}-\frac{9 \sqrt{\frac{5}{7}} \beta ^4 t}{286 s^4}-\frac{3 
\sqrt{\frac{5}{7}} \beta ^4}{572 s^4 t}-\frac{9 \beta ^3 \mu ^3}{11 \sqrt{35} 
s^2}+\frac{9 \beta ^3 \mu }{11 \sqrt{35} s^2}+\frac{\sqrt{\frac{5}{7}} \beta ^3 
\mu ^4 t}{11 s^2}-\frac{\sqrt{\frac{5}{7}} \beta ^3 \mu ^2 t}{22 s^2}+\frac{3 
\beta ^3 \mu ^2}{22 \sqrt{35} s^2 t}-\frac{\sqrt{\frac{5}{7}} \beta ^3 t}{22 
s^2}-\frac{3 \beta ^3}{22 \sqrt{35} s^2 t}+\frac{\beta ^2 \mu ^2}{4 \sqrt{35} 
t}-\frac{\beta ^2}{4 \sqrt{35} t} \Big\rbrace
\label{eq:apt2343}
\end{dmath}
\begin{dmath}
\TB{3}{4}{31} =
\sqrt{1-\mu ^2} \Big\lbrace\frac{6 \sqrt{\frac{5}{7}} \beta ^4 \mu ^4 t^5}{143 
s^4}-\frac{9 \sqrt{\frac{5}{7}} \beta ^4 \mu ^2 t^5}{286 s^4}-\frac{3 
\sqrt{\frac{5}{7}} \beta ^4 t^5}{286 s^4}+\frac{\sqrt{\frac{5}{7}} \beta ^3 \mu 
^4 t^5}{11 s^4}-\frac{\sqrt{\frac{5}{7}} \beta ^3 \mu ^2 t^5}{22 
s^4}-\frac{\sqrt{\frac{5}{7}} \beta ^3 t^5}{22 s^4}-\frac{9 \sqrt{\frac{5}{7}} 
\beta ^4 \mu ^3 t^4}{143 s^4}+\frac{9 \sqrt{\frac{5}{7}} \beta ^4 \mu  t^4}{143 
s^4}-\frac{6 \beta ^3 \mu ^3 t^4}{11 \sqrt{35} s^4}+\frac{6 \beta ^3 \mu  
t^4}{11 \sqrt{35} s^4}+\frac{\beta ^2 \mu ^3 t^4}{\sqrt{35} s^4}-\frac{\beta ^2 
\mu  t^4}{\sqrt{35} s^4}+\frac{9 \sqrt{\frac{5}{7}} \beta ^4 \mu ^2 t^3}{572 
s^4}-\frac{9 \sqrt{\frac{5}{7}} \beta ^4 t^3}{572 s^4}-\frac{3 \beta ^2 \mu ^2 
t^3}{4 \sqrt{35} s^4}+\frac{3 \beta ^2 t^3}{4 \sqrt{35} s^4}+\frac{3 \beta ^3 
\mu ^2 t^3}{22 \sqrt{35} s^2}-\frac{3 \beta ^3 t^3}{22 \sqrt{35} 
s^2}+\frac{\beta ^2 \mu ^2 t^3}{2 \sqrt{35} s^2}-\frac{\beta ^2 t^3}{2 \sqrt{35} 
s^2} \Big\rbrace
\label{eq:apt3143}
\end{dmath}
\begin{dmath}
\TB{4}{4}{12} =
-\frac{3 \beta ^4 \mu ^4}{143 \sqrt{70}}+\frac{3}{143} \sqrt{\frac{2}{35}} \beta 
^4 \mu ^2-\frac{3 \beta ^4}{143 \sqrt{70}}-\frac{3 \beta ^3 \mu ^4}{22 
\sqrt{70}}+\frac{3 \beta ^3 \mu ^2}{11 \sqrt{70}}-\frac{3 \beta ^3}{22 
\sqrt{70}}-\frac{\beta ^2 \mu ^4}{2 \sqrt{70}}+\frac{\beta ^2 \mu 
^2}{\sqrt{70}}-\frac{\beta ^2}{2 \sqrt{70}}+\frac{3}{286} \sqrt{\frac{5}{14}} 
\beta ^4 \mu ^5 t-\frac{3}{143} \sqrt{\frac{5}{14}} \beta ^4 \mu ^3 
t+\frac{3}{286} \sqrt{\frac{5}{14}} \beta ^4 \mu  t+\frac{1}{22} 
\sqrt{\frac{5}{14}} \beta ^3 \mu ^5 t-\frac{1}{11} \sqrt{\frac{5}{14}} \beta ^3 
\mu ^3 t+\frac{1}{22} \sqrt{\frac{5}{14}} \beta ^3 \mu  t
\label{eq:apt1244}
\end{dmath}
\begin{dmath}
\TB{4}{4}{23} =
\frac{9 \beta ^4 \mu ^4}{286 \sqrt{70} s^4}-\frac{9 \beta ^4 \mu ^2}{143 
\sqrt{70} s^4}+\frac{9 \beta ^4}{286 \sqrt{70} s^4}+\frac{\sqrt{\frac{5}{14}} 
\beta ^3 \mu ^5 t^3}{22 s^4}-\frac{\sqrt{\frac{5}{14}} \beta ^3 \mu ^3 t^3}{11 
s^4}+\frac{\sqrt{\frac{5}{14}} \beta ^3 \mu  t^3}{22 s^4}+\frac{3 
\sqrt{\frac{5}{14}} \beta ^4 \mu ^6 t^2}{143 s^4}-\frac{9 \sqrt{\frac{5}{14}} 
\beta ^4 \mu ^4 t^2}{286 s^4}+\frac{3 \sqrt{\frac{5}{14}} \beta ^4 t^2}{286 
s^4}-\frac{\beta ^3 \mu ^4 t^2}{11 \sqrt{70} s^4}+\frac{\sqrt{\frac{2}{35}} 
\beta ^3 \mu ^2 t^2}{11 s^4}-\frac{\beta ^3 t^2}{11 \sqrt{70} s^4}-\frac{9 
\sqrt{\frac{5}{14}} \beta ^4 \mu ^5 t}{286 s^4}+\frac{9 \sqrt{\frac{5}{14}} 
\beta ^4 \mu ^3 t}{143 s^4}-\frac{9 \sqrt{\frac{5}{14}} \beta ^4 \mu  t}{286 
s^4}+\frac{3 \beta ^3 \mu ^4}{22 \sqrt{70} s^2}-\frac{3 \beta ^3 \mu ^2}{11 
\sqrt{70} s^2}+\frac{3 \beta ^3}{22 \sqrt{70} s^2}-\frac{\sqrt{\frac{5}{14}} 
\beta ^3 \mu ^5 t}{22 s^2}+\frac{\sqrt{\frac{5}{14}} \beta ^3 \mu ^3 t}{11 
s^2}-\frac{\sqrt{\frac{5}{14}} \beta ^3 \mu  t}{22 s^2}
\label{eq:apt2344}
\end{dmath}
\begin{dmath}
\TB{4}{4}{31} =
-\frac{3 \sqrt{\frac{5}{14}} \beta ^4 \mu ^5 t^5}{286 s^4}+\frac{3 
\sqrt{\frac{5}{14}} \beta ^4 \mu ^3 t^5}{143 s^4}-\frac{3 \sqrt{\frac{5}{14}} 
\beta ^4 \mu  t^5}{286 s^4}-\frac{\sqrt{\frac{5}{14}} \beta ^3 \mu ^5 t^5}{22 
s^4}+\frac{\sqrt{\frac{5}{14}} \beta ^3 \mu ^3 t^5}{11 
s^4}-\frac{\sqrt{\frac{5}{14}} \beta ^3 \mu  t^5}{22 s^4}+\frac{9 \beta ^4 \mu 
^4 t^4}{286 \sqrt{70} s^4}-\frac{9 \beta ^4 \mu ^2 t^4}{143 \sqrt{70} 
s^4}+\frac{9 \beta ^4 t^4}{286 \sqrt{70} s^4}+\frac{\beta ^3 \mu ^4 t^4}{11 
\sqrt{70} s^4}-\frac{\sqrt{\frac{2}{35}} \beta ^3 \mu ^2 t^4}{11 
s^4}+\frac{\beta ^3 t^4}{11 \sqrt{70} s^4}-\frac{\beta ^2 \mu ^4 t^4}{2 
\sqrt{70} s^4}+\frac{\beta ^2 \mu ^2 t^4}{\sqrt{70} s^4}-\frac{\beta ^2 t^4}{2 
\sqrt{70} s^4}
\label{eq:apt3144}
\end{dmath}
\begin{dmath}
\TB{0}{6}{12} =
-\frac{212 \beta ^4 \mu ^4}{693}+\frac{40 \beta ^4 \mu ^2}{231}+\frac{4 \beta 
^4}{1155}-\frac{5 \beta ^3 \mu ^4}{11}+\frac{18 \beta ^3 \mu ^2}{77}+\frac{\beta 
^3}{77}+\frac{34}{165} \beta ^4 \mu ^5 t-\frac{92}{693} \beta ^4 \mu ^3 
t+\frac{76 \beta ^4 \mu ^3}{693 t}-\frac{2}{231} \beta ^4 \mu  t-\frac{52 \beta 
^4 \mu }{1155 t}+\frac{3}{11} \beta ^3 \mu ^5 t-\frac{10}{77} \beta ^3 \mu ^3 
t+\frac{40 \beta ^3 \mu ^3}{231 t}-\frac{3}{77} \beta ^3 \mu  t-\frac{16 \beta 
^3 \mu }{231 t}
\label{eq:apt1260}
\end{dmath}
\begin{dmath}
\TB{0}{6}{23} =
\frac{106 \beta ^4 \mu ^4}{231 s^4}-\frac{20 \beta ^4 \mu ^2}{77 s^4}-\frac{2 
\beta ^4}{385 s^4}+\frac{8 \beta ^3 \mu ^2}{77 s^4}-\frac{8 \beta ^3}{231 
s^4}+\frac{3 \beta ^3 \mu ^5 t^3}{11 s^4}-\frac{10 \beta ^3 \mu ^3 t^3}{33 
s^4}+\frac{5 \beta ^3 \mu  t^3}{77 s^4}+\frac{83 \beta ^4 \mu ^6 t^2}{330 
s^4}-\frac{3 \beta ^4 \mu ^4 t^2}{22 s^4}-\frac{29 \beta ^4 \mu ^2 t^2}{462 
s^4}+\frac{17 \beta ^4 t^2}{1386 s^4}-\frac{10 \beta ^3 \mu ^4 t^2}{33 
s^4}+\frac{20 \beta ^3 \mu ^2 t^2}{77 s^4}-\frac{2 \beta ^3 t^2}{77 
s^4}-\frac{34 \beta ^4 \mu ^5 t}{55 s^4}+\frac{92 \beta ^4 \mu ^3 t}{231 
s^4}-\frac{76 \beta ^4 \mu ^3}{693 s^4 t}+\frac{2 \beta ^4 \mu  t}{77 
s^4}+\frac{52 \beta ^4 \mu }{1155 s^4 t}-\frac{8 \beta ^3 \mu }{231 s^4 
t}+\frac{5 \beta ^3 \mu ^4}{11 s^2}-\frac{30 \beta ^3 \mu ^2}{77 s^2}+\frac{3 
\beta ^3}{77 s^2}-\frac{3 \beta ^3 \mu ^5 t}{11 s^2}+\frac{10 \beta ^3 \mu ^3 
t}{33 s^2}-\frac{40 \beta ^3 \mu ^3}{231 s^2 t}-\frac{5 \beta ^3 \mu  t}{77 
s^2}+\frac{8 \beta ^3 \mu }{77 s^2 t}
\label{eq:apt2360}
\end{dmath}
\begin{dmath}
\TB{0}{6}{31} =
-\frac{34 \beta ^4 \mu ^5 t^5}{165 s^4}+\frac{92 \beta ^4 \mu ^3 t^5}{693 
s^4}+\frac{2 \beta ^4 \mu  t^5}{231 s^4}-\frac{3 \beta ^3 \mu ^5 t^5}{11 
s^4}+\frac{10 \beta ^3 \mu ^3 t^5}{33 s^4}-\frac{5 \beta ^3 \mu  t^5}{77 
s^4}+\frac{106 \beta ^4 \mu ^4 t^4}{231 s^4}-\frac{20 \beta ^4 \mu ^2 t^4}{77 
s^4}-\frac{2 \beta ^4 t^4}{385 s^4}+\frac{10 \beta ^3 \mu ^4 t^4}{33 
s^4}-\frac{20 \beta ^3 \mu ^2 t^4}{77 s^4}+\frac{2 \beta ^3 t^4}{77 
s^4}-\frac{76 \beta ^4 \mu ^3 t^3}{231 s^4}+\frac{52 \beta ^4 \mu  t^3}{385 
s^4}+\frac{92 \beta ^4 \mu ^2 t^2}{1155 s^4}-\frac{52 \beta ^4 t^2}{3465 
s^4}-\frac{8 \beta ^3 \mu ^2 t^2}{77 s^4}+\frac{8 \beta ^3 t^2}{231 s^4}+\frac{8 
\beta ^3 \mu  t}{231 s^4}-\frac{40 \beta ^3 \mu ^3 t^3}{231 s^2}+\frac{8 \beta 
^3 \mu  t^3}{77 s^2}+\frac{12 \beta ^3 \mu ^2 t^2}{77 s^2}-\frac{4 \beta ^3 
t^2}{77 s^2}-\frac{8 \beta ^3 \mu  t}{231 s^2}
\label{eq:apt3160}
\end{dmath}
\begin{dmath}
\TB{1}{6}{12} =
\sqrt{1-\mu ^2} \Big\lbrace-\frac{136}{165} \sqrt{\frac{2}{21}} \beta ^4 \mu 
^3+\frac{8}{55} \sqrt{\frac{2}{21}} \beta ^4 \mu -\frac{2}{11} 
\sqrt{\frac{14}{3}} \beta ^3 \mu ^3+\frac{2}{11} \sqrt{\frac{2}{21}} \beta ^3 
\mu +\frac{1}{6} \sqrt{\frac{7}{6}} \beta ^4 \mu ^4 t-\frac{1}{11} 
\sqrt{\frac{3}{14}} \beta ^4 \mu ^2 t+\frac{\sqrt{\frac{6}{7}} \beta ^4 \mu 
^2}{11 t}-\frac{\beta ^4 t}{22 \sqrt{42}}-\frac{\sqrt{\frac{2}{21}} \beta ^4}{55 
t}+\frac{5}{22} \sqrt{\frac{7}{6}} \beta ^3 \mu ^4 t-\frac{5 \beta ^3 \mu ^2 
t}{33 \sqrt{42}}+\frac{5 \sqrt{\frac{2}{21}} \beta ^3 \mu ^2}{11 t}-\frac{1}{66} 
\sqrt{\frac{7}{6}} \beta ^3 t-\frac{\sqrt{\frac{2}{21}} \beta ^3}{33 t} 
\Big\rbrace
\label{eq:apt1261}
\end{dmath}
\begin{dmath}
\TB{1}{6}{23} =
\sqrt{1-\mu ^2} \Big\lbrace\frac{68 \sqrt{\frac{2}{21}} \beta ^4 \mu ^3}{55 
s^4}-\frac{4 \sqrt{\frac{6}{7}} \beta ^4 \mu }{55 s^4}+\frac{8 
\sqrt{\frac{2}{21}} \beta ^3 \mu }{33 s^4}+\frac{5 \sqrt{\frac{7}{6}} \beta ^3 
\mu ^4 t^3}{22 s^4}-\frac{5 \sqrt{\frac{7}{6}} \beta ^3 \mu ^2 t^3}{33 
s^4}+\frac{5 \beta ^3 t^3}{66 \sqrt{42} s^4}+\frac{27 \sqrt{\frac{3}{14}} \beta 
^4 \mu ^5 t^2}{55 s^4}-\frac{\sqrt{\frac{2}{21}} \beta ^4 \mu ^3 t^2}{11 
s^4}-\frac{\sqrt{\frac{3}{14}} \beta ^4 \mu  t^2}{11 s^4}-\frac{4 
\sqrt{\frac{14}{3}} \beta ^3 \mu ^3 t^2}{33 s^4}+\frac{4 \sqrt{\frac{2}{21}} 
\beta ^3 \mu  t^2}{11 s^4}-\frac{\sqrt{\frac{7}{6}} \beta ^4 \mu ^4 t}{2 
s^4}+\frac{3 \sqrt{\frac{3}{14}} \beta ^4 \mu ^2 t}{11 
s^4}-\frac{\sqrt{\frac{6}{7}} \beta ^4 \mu ^2}{11 s^4 
t}+\frac{\sqrt{\frac{3}{14}} \beta ^4 t}{22 s^4}+\frac{\sqrt{\frac{2}{21}} \beta 
^4}{55 s^4 t}-\frac{2 \sqrt{\frac{2}{21}} \beta ^3}{33 s^4 t}+\frac{2 
\sqrt{\frac{14}{3}} \beta ^3 \mu ^3}{11 s^2}-\frac{2 \sqrt{\frac{6}{7}} \beta ^3 
\mu }{11 s^2}-\frac{5 \sqrt{\frac{7}{6}} \beta ^3 \mu ^4 t}{22 s^2}+\frac{5 
\sqrt{\frac{7}{6}} \beta ^3 \mu ^2 t}{33 s^2}-\frac{5 \sqrt{\frac{2}{21}} \beta 
^3 \mu ^2}{11 s^2 t}-\frac{5 \beta ^3 t}{66 \sqrt{42} 
s^2}+\frac{\sqrt{\frac{2}{21}} \beta ^3}{11 s^2 t} \Big\rbrace
\label{eq:apt2361}
\end{dmath}
\begin{dmath}
\TB{1}{6}{31} =
\sqrt{1-\mu ^2} \Big\lbrace-\frac{\sqrt{\frac{7}{6}} \beta ^4 \mu ^4 t^5}{6 
s^4}+\frac{\sqrt{\frac{3}{14}} \beta ^4 \mu ^2 t^5}{11 s^4}+\frac{\beta ^4 
t^5}{22 \sqrt{42} s^4}-\frac{5 \sqrt{\frac{7}{6}} \beta ^3 \mu ^4 t^5}{22 
s^4}+\frac{5 \sqrt{\frac{7}{6}} \beta ^3 \mu ^2 t^5}{33 s^4}-\frac{5 \beta ^3 
t^5}{66 \sqrt{42} s^4}+\frac{68 \sqrt{\frac{2}{21}} \beta ^4 \mu ^3 t^4}{55 
s^4}-\frac{4 \sqrt{\frac{6}{7}} \beta ^4 \mu  t^4}{55 s^4}+\frac{4 
\sqrt{\frac{14}{3}} \beta ^3 \mu ^3 t^4}{33 s^4}-\frac{4 \sqrt{\frac{2}{21}} 
\beta ^3 \mu  t^4}{11 s^4}-\frac{3 \sqrt{\frac{6}{7}} \beta ^4 \mu ^2 t^3}{11 
s^4}+\frac{\sqrt{\frac{6}{7}} \beta ^4 t^3}{55 s^4}+\frac{4 \sqrt{\frac{14}{3}} 
\beta ^4 \mu  t^2}{165 s^4}-\frac{8 \sqrt{\frac{2}{21}} \beta ^3 \mu  t^2}{33 
s^4}+\frac{2 \sqrt{\frac{2}{21}} \beta ^3 t}{33 s^4}-\frac{5 \sqrt{\frac{2}{21}} 
\beta ^3 \mu ^2 t^3}{11 s^2}+\frac{\sqrt{\frac{2}{21}} \beta ^3 t^3}{11 
s^2}+\frac{4 \sqrt{\frac{2}{21}} \beta ^3 \mu  t^2}{11 s^2}-\frac{2 
\sqrt{\frac{2}{21}} \beta ^3 t}{33 s^2} \Big\rbrace
\label{eq:apt3161}
\end{dmath}
\begin{dmath}
\TB{2}{6}{12} =
\frac{16 \beta ^4 \mu ^4}{11 \sqrt{105}}-\frac{16 \beta ^4 \mu ^2}{11 
\sqrt{105}}+\frac{4}{11} \sqrt{\frac{7}{15}} \beta ^3 \mu ^4-\frac{4}{11} 
\sqrt{\frac{7}{15}} \beta ^3 \mu ^2-\frac{8}{33} \sqrt{\frac{5}{21}} \beta ^4 
\mu ^5 t+\frac{8}{33} \sqrt{\frac{5}{21}} \beta ^4 \mu ^3 t-\frac{4 \beta ^4 \mu 
^3}{11 \sqrt{105} t}+\frac{4 \beta ^4 \mu }{11 \sqrt{105} t}-\frac{4}{11} 
\sqrt{\frac{5}{21}} \beta ^3 \mu ^5 t+\frac{8}{33} \sqrt{\frac{7}{15}} \beta ^3 
\mu ^3 t-\frac{8 \beta ^3 \mu ^3}{11 \sqrt{105} t}+\frac{4 \beta ^3 \mu  t}{33 
\sqrt{105}}+\frac{8 \beta ^3 \mu }{11 \sqrt{105} t}
\label{eq:apt1262}
\end{dmath}
\begin{dmath}
\TB{2}{6}{23} =
-\frac{8 \sqrt{\frac{3}{35}} \beta ^4 \mu ^4}{11 s^4}+\frac{8 
\sqrt{\frac{3}{35}} \beta ^4 \mu ^2}{11 s^4}-\frac{8 \beta ^3 \mu ^2}{33 
\sqrt{105} s^4}+\frac{8 \beta ^3}{33 \sqrt{105} s^4}-\frac{4 \sqrt{\frac{5}{21}} 
\beta ^3 \mu ^5 t^3}{11 s^4}+\frac{16 \sqrt{\frac{5}{21}} \beta ^3 \mu ^3 
t^3}{33 s^4}-\frac{4 \sqrt{\frac{5}{21}} \beta ^3 \mu  t^3}{33 s^4}-\frac{5 
\sqrt{\frac{15}{7}} \beta ^4 \mu ^6 t^2}{44 s^4}+\frac{13 \sqrt{\frac{5}{21}} 
\beta ^4 \mu ^4 t^2}{44 s^4}+\frac{\sqrt{\frac{15}{7}} \beta ^4 \mu ^2 t^2}{44 
s^4}-\frac{\sqrt{\frac{5}{21}} \beta ^4 t^2}{44 s^4}+\frac{8 \sqrt{\frac{7}{15}} 
\beta ^3 \mu ^4 t^2}{33 s^4}-\frac{64 \beta ^3 \mu ^2 t^2}{33 \sqrt{105} 
s^4}+\frac{8 \beta ^3 t^2}{33 \sqrt{105} s^4}+\frac{8 \sqrt{\frac{5}{21}} \beta 
^4 \mu ^5 t}{11 s^4}-\frac{8 \sqrt{\frac{5}{21}} \beta ^4 \mu ^3 t}{11 
s^4}+\frac{4 \beta ^4 \mu ^3}{11 \sqrt{105} s^4 t}-\frac{4 \beta ^4 \mu }{11 
\sqrt{105} s^4 t}-\frac{4 \sqrt{\frac{7}{15}} \beta ^3 \mu ^4}{11 s^2}+\frac{32 
\beta ^3 \mu ^2}{11 \sqrt{105} s^2}-\frac{4 \beta ^3}{11 \sqrt{105} s^2}+\frac{4 
\sqrt{\frac{5}{21}} \beta ^3 \mu ^5 t}{11 s^2}-\frac{16 \sqrt{\frac{5}{21}} 
\beta ^3 \mu ^3 t}{33 s^2}+\frac{8 \beta ^3 \mu ^3}{11 \sqrt{105} s^2 t}+\frac{4 
\sqrt{\frac{5}{21}} \beta ^3 \mu  t}{33 s^2}-\frac{8 \beta ^3 \mu }{11 
\sqrt{105} s^2 t}
\label{eq:apt2362}
\end{dmath}
\begin{dmath}
\TB{2}{6}{31} =
\frac{8 \sqrt{\frac{5}{21}} \beta ^4 \mu ^5 t^5}{33 s^4}-\frac{8 
\sqrt{\frac{5}{21}} \beta ^4 \mu ^3 t^5}{33 s^4}+\frac{4 \sqrt{\frac{5}{21}} 
\beta ^3 \mu ^5 t^5}{11 s^4}-\frac{16 \sqrt{\frac{5}{21}} \beta ^3 \mu ^3 
t^5}{33 s^4}+\frac{4 \sqrt{\frac{5}{21}} \beta ^3 \mu  t^5}{33 s^4}-\frac{8 
\sqrt{\frac{3}{35}} \beta ^4 \mu ^4 t^4}{11 s^4}+\frac{8 \sqrt{\frac{3}{35}} 
\beta ^4 \mu ^2 t^4}{11 s^4}-\frac{8 \sqrt{\frac{7}{15}} \beta ^3 \mu ^4 t^4}{33 
s^4}+\frac{64 \beta ^3 \mu ^2 t^4}{33 \sqrt{105} s^4}-\frac{8 \beta ^3 t^4}{33 
\sqrt{105} s^4}+\frac{4 \sqrt{\frac{3}{35}} \beta ^4 \mu ^3 t^3}{11 s^4}-\frac{4 
\sqrt{\frac{3}{35}} \beta ^4 \mu  t^3}{11 s^4}-\frac{4 \beta ^4 \mu ^2 t^2}{33 
\sqrt{105} s^4}+\frac{4 \beta ^4 t^2}{33 \sqrt{105} s^4}+\frac{8 \beta ^3 \mu ^2 
t^2}{33 \sqrt{105} s^4}-\frac{8 \beta ^3 t^2}{33 \sqrt{105} s^4}+\frac{8 \beta 
^3 \mu ^3 t^3}{11 \sqrt{105} s^2}-\frac{8 \beta ^3 \mu  t^3}{11 \sqrt{105} 
s^2}-\frac{4 \beta ^3 \mu ^2 t^2}{11 \sqrt{105} s^2}+\frac{4 \beta ^3 t^2}{11 
\sqrt{105} s^2}
\label{eq:apt3162}
\end{dmath}
\begin{dmath}
\TB{3}{6}{12} =
\sqrt{1-\mu ^2} \Big\lbrace\frac{16 \beta ^4 \mu ^3}{33 \sqrt{105}}-\frac{16 
\beta ^4 \mu }{33 \sqrt{105}}+\frac{4}{11} \sqrt{\frac{3}{35}} \beta ^3 \mu 
^3-\frac{4}{11} \sqrt{\frac{3}{35}} \beta ^3 \mu -\frac{5}{44} 
\sqrt{\frac{5}{21}} \beta ^4 \mu ^4 t+\frac{1}{66} \sqrt{\frac{35}{3}} \beta ^4 
\mu ^2 t-\frac{2 \beta ^4 \mu ^2}{33 \sqrt{105} t}+\frac{1}{132} 
\sqrt{\frac{5}{21}} \beta ^4 t+\frac{2 \beta ^4}{33 \sqrt{105} t}-\frac{3}{44} 
\sqrt{\frac{15}{7}} \beta ^3 \mu ^4 t+\frac{1}{22} \sqrt{\frac{21}{5}} \beta ^3 
\mu ^2 t-\frac{2 \beta ^3 \mu ^2}{11 \sqrt{105} t}+\frac{1}{44} 
\sqrt{\frac{3}{35}} \beta ^3 t+\frac{2 \beta ^3}{11 \sqrt{105} t} \Big\rbrace
\label{eq:apt1263}
\end{dmath}
\begin{dmath}
\TB{3}{6}{23} =
\sqrt{1-\mu ^2} \Big\lbrace-\frac{8 \beta ^4 \mu ^3}{11 \sqrt{105} s^4}+\frac{8 
\beta ^4 \mu }{11 \sqrt{105} s^4}-\frac{3 \sqrt{\frac{15}{7}} \beta ^3 \mu ^4 
t^3}{44 s^4}+\frac{5 \sqrt{\frac{5}{21}} \beta ^3 \mu ^2 t^3}{22 
s^4}-\frac{\sqrt{\frac{5}{21}} \beta ^3 t^3}{44 s^4}-\frac{13 
\sqrt{\frac{5}{21}} \beta ^4 \mu ^5 t^2}{66 s^4}+\frac{5 \sqrt{\frac{5}{21}} 
\beta ^4 \mu ^3 t^2}{33 s^4}+\frac{\sqrt{\frac{5}{21}} \beta ^4 \mu  t^2}{22 
s^4}+\frac{8 \beta ^3 \mu ^3 t^2}{11 \sqrt{105} s^4}-\frac{8 \beta ^3 \mu  
t^2}{11 \sqrt{105} s^4}+\frac{5 \sqrt{\frac{15}{7}} \beta ^4 \mu ^4 t}{44 
s^4}-\frac{\sqrt{\frac{35}{3}} \beta ^4 \mu ^2 t}{22 s^4}+\frac{2 \beta ^4 \mu 
^2}{33 \sqrt{105} s^4 t}-\frac{\sqrt{\frac{5}{21}} \beta ^4 t}{44 s^4}-\frac{2 
\beta ^4}{33 \sqrt{105} s^4 t}-\frac{4 \sqrt{\frac{3}{35}} \beta ^3 \mu ^3}{11 
s^2}+\frac{4 \sqrt{\frac{3}{35}} \beta ^3 \mu }{11 s^2}+\frac{3 
\sqrt{\frac{15}{7}} \beta ^3 \mu ^4 t}{44 s^2}-\frac{5 \sqrt{\frac{5}{21}} \beta 
^3 \mu ^2 t}{22 s^2}+\frac{2 \beta ^3 \mu ^2}{11 \sqrt{105} s^2 
t}+\frac{\sqrt{\frac{5}{21}} \beta ^3 t}{44 s^2}-\frac{2 \beta ^3}{11 \sqrt{105} 
s^2 t} \Big\rbrace
\label{eq:apt2363}
\end{dmath}
\begin{dmath}
\TB{3}{6}{31} =
\sqrt{1-\mu ^2} \Big\lbrace\frac{5 \sqrt{\frac{5}{21}} \beta ^4 \mu ^4 t^5}{44 
s^4}-\frac{\sqrt{\frac{35}{3}} \beta ^4 \mu ^2 t^5}{66 
s^4}-\frac{\sqrt{\frac{5}{21}} \beta ^4 t^5}{132 s^4}+\frac{3 
\sqrt{\frac{15}{7}} \beta ^3 \mu ^4 t^5}{44 s^4}-\frac{5 \sqrt{\frac{5}{21}} 
\beta ^3 \mu ^2 t^5}{22 s^4}+\frac{\sqrt{\frac{5}{21}} \beta ^3 t^5}{44 
s^4}-\frac{8 \beta ^4 \mu ^3 t^4}{11 \sqrt{105} s^4}+\frac{8 \beta ^4 \mu  
t^4}{11 \sqrt{105} s^4}-\frac{8 \beta ^3 \mu ^3 t^4}{11 \sqrt{105} s^4}+\frac{8 
\beta ^3 \mu  t^4}{11 \sqrt{105} s^4}+\frac{2 \beta ^4 \mu ^2 t^3}{11 \sqrt{105} 
s^4}-\frac{2 \beta ^4 t^3}{11 \sqrt{105} s^4}+\frac{2 \beta ^3 \mu ^2 t^3}{11 
\sqrt{105} s^2}-\frac{2 \beta ^3 t^3}{11 \sqrt{105} s^2} \Big\rbrace
\label{eq:apt3163}
\end{dmath}
\begin{dmath}
\TB{4}{6}{12} =
-\frac{2}{165} \sqrt{\frac{2}{7}} \beta ^4 \mu ^4+\frac{4}{165} 
\sqrt{\frac{2}{7}} \beta ^4 \mu ^2-\frac{2}{165} \sqrt{\frac{2}{7}} \beta 
^4-\frac{\beta ^3 \mu ^4}{11 \sqrt{14}}+\frac{1}{11} \sqrt{\frac{2}{7}} \beta ^3 
\mu ^2-\frac{\beta ^3}{11 \sqrt{14}}+\frac{1}{33} \sqrt{\frac{2}{7}} \beta ^4 
\mu ^5 t-\frac{2}{33} \sqrt{\frac{2}{7}} \beta ^4 \mu ^3 t+\frac{1}{33} 
\sqrt{\frac{2}{7}} \beta ^4 \mu  t+\frac{5 \beta ^3 \mu ^5 t}{33 
\sqrt{14}}-\frac{5}{33} \sqrt{\frac{2}{7}} \beta ^3 \mu ^3 t+\frac{5 \beta ^3 
\mu  t}{33 \sqrt{14}}
\label{eq:apt1264}
\end{dmath}
\begin{dmath}
\TB{4}{6}{23} =
\frac{\sqrt{\frac{2}{7}} \beta ^4 \mu ^4}{55 s^4}-\frac{2 \sqrt{\frac{2}{7}} 
\beta ^4 \mu ^2}{55 s^4}+\frac{\sqrt{\frac{2}{7}} \beta ^4}{55 s^4}+\frac{5 
\beta ^3 \mu ^5 t^3}{33 \sqrt{14} s^4}-\frac{5 \sqrt{\frac{2}{7}} \beta ^3 \mu 
^3 t^3}{33 s^4}+\frac{5 \beta ^3 \mu  t^3}{33 \sqrt{14} s^4}+\frac{17 \beta ^4 
\mu ^6 t^2}{110 \sqrt{14} s^4}-\frac{31 \beta ^4 \mu ^4 t^2}{110 \sqrt{14} 
s^4}+\frac{\beta ^4 \mu ^2 t^2}{10 \sqrt{14} s^4}+\frac{3 \beta ^4 t^2}{110 
\sqrt{14} s^4}-\frac{\sqrt{\frac{2}{7}} \beta ^3 \mu ^4 t^2}{33 s^4}+\frac{2 
\sqrt{\frac{2}{7}} \beta ^3 \mu ^2 t^2}{33 s^4}-\frac{\sqrt{\frac{2}{7}} \beta 
^3 t^2}{33 s^4}-\frac{\sqrt{\frac{2}{7}} \beta ^4 \mu ^5 t}{11 s^4}+\frac{2 
\sqrt{\frac{2}{7}} \beta ^4 \mu ^3 t}{11 s^4}-\frac{\sqrt{\frac{2}{7}} \beta ^4 
\mu  t}{11 s^4}+\frac{\beta ^3 \mu ^4}{11 \sqrt{14} 
s^2}-\frac{\sqrt{\frac{2}{7}} \beta ^3 \mu ^2}{11 s^2}+\frac{\beta ^3}{11 
\sqrt{14} s^2}-\frac{5 \beta ^3 \mu ^5 t}{33 \sqrt{14} s^2}+\frac{5 
\sqrt{\frac{2}{7}} \beta ^3 \mu ^3 t}{33 s^2}-\frac{5 \beta ^3 \mu  t}{33 
\sqrt{14} s^2}
\label{eq:apt2364}
\end{dmath}
\begin{dmath}
\TB{4}{6}{31} =
-\frac{\sqrt{\frac{2}{7}} \beta ^4 \mu ^5 t^5}{33 s^4}+\frac{2 
\sqrt{\frac{2}{7}} \beta ^4 \mu ^3 t^5}{33 s^4}-\frac{\sqrt{\frac{2}{7}} \beta 
^4 \mu  t^5}{33 s^4}-\frac{5 \beta ^3 \mu ^5 t^5}{33 \sqrt{14} s^4}+\frac{5 
\sqrt{\frac{2}{7}} \beta ^3 \mu ^3 t^5}{33 s^4}-\frac{5 \beta ^3 \mu  t^5}{33 
\sqrt{14} s^4}+\frac{\sqrt{\frac{2}{7}} \beta ^4 \mu ^4 t^4}{55 s^4}-\frac{2 
\sqrt{\frac{2}{7}} \beta ^4 \mu ^2 t^4}{55 s^4}+\frac{\sqrt{\frac{2}{7}} \beta 
^4 t^4}{55 s^4}+\frac{\sqrt{\frac{2}{7}} \beta ^3 \mu ^4 t^4}{33 s^4}-\frac{2 
\sqrt{\frac{2}{7}} \beta ^3 \mu ^2 t^4}{33 s^4}+\frac{\sqrt{\frac{2}{7}} \beta 
^3 t^4}{33 s^4}
\label{eq:apt3164}
\end{dmath}
\begin{dmath}
\TB{5}{6}{12} =
\sqrt{1-\mu ^2} \Big\lbrace\frac{\beta ^4 \mu ^4 t}{60 \sqrt{77}}-\frac{\beta ^4 
\mu ^2 t}{30 \sqrt{77}}+\frac{\beta ^4 t}{60 \sqrt{77}}+\frac{\beta ^3 \mu ^4 
t}{12 \sqrt{77}}-\frac{\beta ^3 \mu ^2 t}{6 \sqrt{77}}+\frac{\beta ^3 t}{12 
\sqrt{77}} \Big\rbrace
\label{eq:apt1265}
\end{dmath}
\begin{dmath}
\TB{5}{6}{23} =
\sqrt{1-\mu ^2} \Big\lbrace\frac{\beta ^3 \mu ^4 t^3}{12 \sqrt{77} 
s^4}-\frac{\beta ^3 \mu ^2 t^3}{6 \sqrt{77} s^4}+\frac{\beta ^3 t^3}{12 
\sqrt{77} s^4}+\frac{\beta ^4 \mu ^5 t^2}{10 \sqrt{77} s^4}-\frac{\beta ^4 \mu 
^3 t^2}{5 \sqrt{77} s^4}+\frac{\beta ^4 \mu  t^2}{10 \sqrt{77} s^4}-\frac{\beta 
^4 \mu ^4 t}{20 \sqrt{77} s^4}+\frac{\beta ^4 \mu ^2 t}{10 \sqrt{77} 
s^4}-\frac{\beta ^4 t}{20 \sqrt{77} s^4}-\frac{\beta ^3 \mu ^4 t}{12 \sqrt{77} 
s^2}+\frac{\beta ^3 \mu ^2 t}{6 \sqrt{77} s^2}-\frac{\beta ^3 t}{12 \sqrt{77} 
s^2} \Big\rbrace
\label{eq:apt2365}
\end{dmath}
\begin{dmath}
\TB{5}{6}{31} =
\sqrt{1-\mu ^2} \Big\lbrace-\frac{\beta ^4 \mu ^4 t^5}{60 \sqrt{77} 
s^4}+\frac{\beta ^4 \mu ^2 t^5}{30 \sqrt{77} s^4}-\frac{\beta ^4 t^5}{60 
\sqrt{77} s^4}-\frac{\beta ^3 \mu ^4 t^5}{12 \sqrt{77} s^4}+\frac{\beta ^3 \mu 
^2 t^5}{6 \sqrt{77} s^4}-\frac{\beta ^3 t^5}{12 \sqrt{77} s^4} \Big\rbrace
\label{eq:apt3165}
\end{dmath}
\begin{dmath}
\TB{6}{6}{23} =
-\frac{\beta ^4 \mu ^6 t^2}{60 \sqrt{231} s^4}+\frac{\beta ^4 \mu ^4 t^2}{20 
\sqrt{231} s^4}-\frac{\beta ^4 \mu ^2 t^2}{20 \sqrt{231} s^4}+\frac{\beta ^4 
t^2}{60 \sqrt{231} s^4}
\label{eq:apt2366}
\end{dmath}
\begin{dmath}
\TB{0}{8}{12} =
-\frac{112 \beta ^4 \mu ^4}{1287}+\frac{32 \beta ^4 \mu ^2}{429}-\frac{16 \beta 
^4}{2145}+\frac{56}{715} \beta ^4 \mu ^5 t-\frac{112 \beta ^4 \mu ^3 
t}{1287}+\frac{32 \beta ^4 \mu ^3}{1287 t}+\frac{8}{429} \beta ^4 \mu  
t-\frac{32 \beta ^4 \mu }{2145 t}
\label{eq:apt1280}
\end{dmath}
\begin{dmath}
\TB{0}{8}{23} =
\frac{56 \beta ^4 \mu ^4}{429 s^4}-\frac{16 \beta ^4 \mu ^2}{143 s^4}+\frac{8 
\beta ^4}{715 s^4}+\frac{28 \beta ^4 \mu ^6 t^2}{195 s^4}-\frac{28 \beta ^4 \mu 
^4 t^2}{143 s^4}+\frac{28 \beta ^4 \mu ^2 t^2}{429 s^4}-\frac{4 \beta ^4 
t^2}{1287 s^4}-\frac{168 \beta ^4 \mu ^5 t}{715 s^4}+\frac{112 \beta ^4 \mu ^3 
t}{429 s^4}-\frac{32 \beta ^4 \mu ^3}{1287 s^4 t}-\frac{8 \beta ^4 \mu  t}{143 
s^4}+\frac{32 \beta ^4 \mu }{2145 s^4 t}
\label{eq:apt2380}
\end{dmath}
\begin{dmath}
\TB{0}{8}{31} =
-\frac{56 \beta ^4 \mu ^5 t^5}{715 s^4}+\frac{112 \beta ^4 \mu ^3 t^5}{1287 
s^4}-\frac{8 \beta ^4 \mu  t^5}{429 s^4}+\frac{56 \beta ^4 \mu ^4 t^4}{429 
s^4}-\frac{16 \beta ^4 \mu ^2 t^4}{143 s^4}+\frac{8 \beta ^4 t^4}{715 
s^4}-\frac{32 \beta ^4 \mu ^3 t^3}{429 s^4}+\frac{32 \beta ^4 \mu  t^3}{715 
s^4}+\frac{32 \beta ^4 \mu ^2 t^2}{2145 s^4}-\frac{32 \beta ^4 t^2}{6435 s^4}
\label{eq:apt3180}
\end{dmath}
\begin{dmath}
\TB{1}{8}{12} =
\sqrt{1-\mu ^2} \Big\lbrace-\frac{112 \sqrt{2} \beta ^4 \mu 
^3}{2145}+\frac{16}{715} \sqrt{2} \beta ^4 \mu +\frac{7}{143} \sqrt{2} \beta ^4 
\mu ^4 t-\frac{14}{429} \sqrt{2} \beta ^4 \mu ^2 t+\frac{2 \sqrt{2} \beta ^4 \mu 
^2}{143 t}+\frac{1}{429} \sqrt{2} \beta ^4 t-\frac{2 \sqrt{2} \beta ^4}{715 t} 
\Big\rbrace
\label{eq:apt1281}
\end{dmath}
\begin{dmath}
\TB{1}{8}{23} =
\sqrt{1-\mu ^2} \Big\lbrace\frac{56 \sqrt{2} \beta ^4 \mu ^3}{715 s^4}-\frac{24 
\sqrt{2} \beta ^4 \mu }{715 s^4}+\frac{6 \sqrt{2} \beta ^4 \mu ^5 t^2}{65 
s^4}-\frac{12 \sqrt{2} \beta ^4 \mu ^3 t^2}{143 s^4}+\frac{2 \sqrt{2} \beta ^4 
\mu  t^2}{143 s^4}-\frac{21 \sqrt{2} \beta ^4 \mu ^4 t}{143 s^4}+\frac{14 
\sqrt{2} \beta ^4 \mu ^2 t}{143 s^4}-\frac{2 \sqrt{2} \beta ^4 \mu ^2}{143 s^4 
t}-\frac{\sqrt{2} \beta ^4 t}{143 s^4}+\frac{2 \sqrt{2} \beta ^4}{715 s^4 t} 
\Big\rbrace
\label{eq:apt2381}
\end{dmath}
\begin{dmath}
\TB{1}{8}{31} =
\sqrt{1-\mu ^2} \Big\lbrace-\frac{7 \sqrt{2} \beta ^4 \mu ^4 t^5}{143 
s^4}+\frac{14 \sqrt{2} \beta ^4 \mu ^2 t^5}{429 s^4}-\frac{\sqrt{2} \beta ^4 
t^5}{429 s^4}+\frac{56 \sqrt{2} \beta ^4 \mu ^3 t^4}{715 s^4}-\frac{24 \sqrt{2} 
\beta ^4 \mu  t^4}{715 s^4}-\frac{6 \sqrt{2} \beta ^4 \mu ^2 t^3}{143 
s^4}+\frac{6 \sqrt{2} \beta ^4 t^3}{715 s^4}+\frac{16 \sqrt{2} \beta ^4 \mu  
t^2}{2145 s^4} \Big\rbrace
\label{eq:apt3181}
\end{dmath}
\begin{dmath}
\TB{2}{8}{12} =
\frac{16}{429} \sqrt{\frac{7}{5}} \beta ^4 \mu ^4-\frac{128 \beta ^4 \mu ^2}{429 
\sqrt{35}}+\frac{16 \beta ^4}{429 \sqrt{35}}-\frac{8}{143} \sqrt{\frac{5}{7}} 
\beta ^4 \mu ^5 t+\frac{32}{429} \sqrt{\frac{5}{7}} \beta ^4 \mu ^3 t-\frac{8 
\beta ^4 \mu ^3}{143 \sqrt{35} t}-\frac{8}{429} \sqrt{\frac{5}{7}} \beta ^4 \mu  
t+\frac{8 \beta ^4 \mu }{143 \sqrt{35} t}
\label{eq:apt1282}
\end{dmath}
\begin{dmath}
\TB{2}{8}{23} =
-\frac{8 \sqrt{\frac{7}{5}} \beta ^4 \mu ^4}{143 s^4}+\frac{64 \beta ^4 \mu 
^2}{143 \sqrt{35} s^4}-\frac{8 \beta ^4}{143 \sqrt{35} s^4}-\frac{3 
\sqrt{\frac{5}{7}} \beta ^4 \mu ^6 t^2}{26 s^4}+\frac{51 \sqrt{\frac{5}{7}} 
\beta ^4 \mu ^4 t^2}{286 s^4}-\frac{19 \sqrt{\frac{5}{7}} \beta ^4 \mu ^2 
t^2}{286 s^4}+\frac{\sqrt{\frac{5}{7}} \beta ^4 t^2}{286 s^4}+\frac{24 
\sqrt{\frac{5}{7}} \beta ^4 \mu ^5 t}{143 s^4}-\frac{32 \sqrt{\frac{5}{7}} \beta 
^4 \mu ^3 t}{143 s^4}+\frac{8 \beta ^4 \mu ^3}{143 \sqrt{35} s^4 t}+\frac{8 
\sqrt{\frac{5}{7}} \beta ^4 \mu  t}{143 s^4}-\frac{8 \beta ^4 \mu }{143 
\sqrt{35} s^4 t}
\label{eq:apt2382}
\end{dmath}
\begin{dmath}
\TB{2}{8}{31} =
\frac{8 \sqrt{\frac{5}{7}} \beta ^4 \mu ^5 t^5}{143 s^4}-\frac{32 
\sqrt{\frac{5}{7}} \beta ^4 \mu ^3 t^5}{429 s^4}+\frac{8 \sqrt{\frac{5}{7}} 
\beta ^4 \mu  t^5}{429 s^4}-\frac{8 \sqrt{\frac{7}{5}} \beta ^4 \mu ^4 t^4}{143 
s^4}+\frac{64 \beta ^4 \mu ^2 t^4}{143 \sqrt{35} s^4}-\frac{8 \beta ^4 t^4}{143 
\sqrt{35} s^4}+\frac{24 \beta ^4 \mu ^3 t^3}{143 \sqrt{35} s^4}-\frac{24 \beta 
^4 \mu  t^3}{143 \sqrt{35} s^4}-\frac{8 \beta ^4 \mu ^2 t^2}{429 \sqrt{35} 
s^4}+\frac{8 \beta ^4 t^2}{429 \sqrt{35} s^4}
\label{eq:apt3182}
\end{dmath}
\begin{dmath}
\TB{3}{8}{12} =
\sqrt{1-\mu ^2} \Big\lbrace\frac{16}{39} \sqrt{\frac{2}{1155}} \beta ^4 \mu 
^3-\frac{16}{39} \sqrt{\frac{2}{1155}} \beta ^4 \mu -\frac{1}{13} 
\sqrt{\frac{15}{154}} \beta ^4 \mu ^4 t+\frac{5}{39} \sqrt{\frac{10}{231}} \beta 
^4 \mu ^2 t-\frac{2 \sqrt{\frac{2}{1155}} \beta ^4 \mu ^2}{39 t}-\frac{1}{39} 
\sqrt{\frac{5}{462}} \beta ^4 t+\frac{2 \sqrt{\frac{2}{1155}} \beta ^4}{39 t} 
\Big\rbrace
\label{eq:apt1283}
\end{dmath}
\begin{dmath}
\TB{3}{8}{23} =
\sqrt{1-\mu ^2} \Big\lbrace-\frac{8 \sqrt{\frac{2}{1155}} \beta ^4 \mu ^3}{13 
s^4}+\frac{8 \sqrt{\frac{2}{1155}} \beta ^4 \mu }{13 
s^4}-\frac{\sqrt{\frac{110}{21}} \beta ^4 \mu ^5 t^2}{39 s^4}+\frac{2 
\sqrt{\frac{70}{33}} \beta ^4 \mu ^3 t^2}{39 s^4}-\frac{\sqrt{\frac{10}{231}} 
\beta ^4 \mu  t^2}{13 s^4}+\frac{3 \sqrt{\frac{15}{154}} \beta ^4 \mu ^4 t}{13 
s^4}-\frac{5 \sqrt{\frac{10}{231}} \beta ^4 \mu ^2 t}{13 s^4}+\frac{2 
\sqrt{\frac{2}{1155}} \beta ^4 \mu ^2}{39 s^4 t}+\frac{\sqrt{\frac{5}{462}} 
\beta ^4 t}{13 s^4}-\frac{2 \sqrt{\frac{2}{1155}} \beta ^4}{39 s^4 t} 
\Big\rbrace
\label{eq:apt2383}
\end{dmath}
\begin{dmath}
\TB{3}{8}{31} =
\sqrt{1-\mu ^2} \Big\lbrace\frac{\sqrt{\frac{15}{154}} \beta ^4 \mu ^4 t^5}{13 
s^4}-\frac{5 \sqrt{\frac{10}{231}} \beta ^4 \mu ^2 t^5}{39 
s^4}+\frac{\sqrt{\frac{5}{462}} \beta ^4 t^5}{39 s^4}-\frac{8 
\sqrt{\frac{2}{1155}} \beta ^4 \mu ^3 t^4}{13 s^4}+\frac{8 \sqrt{\frac{2}{1155}} 
\beta ^4 \mu  t^4}{13 s^4}+\frac{2 \sqrt{\frac{2}{1155}} \beta ^4 \mu ^2 t^3}{13 
s^4}-\frac{2 \sqrt{\frac{2}{1155}} \beta ^4 t^3}{13 s^4} \Big\rbrace
\label{eq:apt3183}
\end{dmath}
\begin{dmath}
\TB{4}{8}{12} =
-\frac{4}{195} \sqrt{\frac{2}{77}} \beta ^4 \mu ^4+\frac{8}{195} 
\sqrt{\frac{2}{77}} \beta ^4 \mu ^2-\frac{4}{195} \sqrt{\frac{2}{77}} \beta 
^4+\frac{2}{39} \sqrt{\frac{2}{77}} \beta ^4 \mu ^5 t-\frac{4}{39} 
\sqrt{\frac{2}{77}} \beta ^4 \mu ^3 t+\frac{2}{39} \sqrt{\frac{2}{77}} \beta ^4 
\mu  t
\label{eq:apt1284}
\end{dmath}
\begin{dmath}
\TB{4}{8}{23} =
\frac{2 \sqrt{\frac{2}{77}} \beta ^4 \mu ^4}{65 s^4}-\frac{4 \sqrt{\frac{2}{77}} 
\beta ^4 \mu ^2}{65 s^4}+\frac{2 \sqrt{\frac{2}{77}} \beta ^4}{65 
s^4}+\frac{\sqrt{\frac{22}{7}} \beta ^4 \mu ^6 t^2}{65 s^4}-\frac{23 
\sqrt{\frac{2}{77}} \beta ^4 \mu ^4 t^2}{65 s^4}+\frac{\sqrt{\frac{2}{77}} \beta 
^4 \mu ^2 t^2}{5 s^4}-\frac{\sqrt{\frac{2}{77}} \beta ^4 t^2}{65 s^4}-\frac{2 
\sqrt{\frac{2}{77}} \beta ^4 \mu ^5 t}{13 s^4}+\frac{4 \sqrt{\frac{2}{77}} \beta 
^4 \mu ^3 t}{13 s^4}-\frac{2 \sqrt{\frac{2}{77}} \beta ^4 \mu  t}{13 s^4}
\label{eq:apt2384}
\end{dmath}
\begin{dmath}
\TB{4}{8}{31} =
-\frac{2 \sqrt{\frac{2}{77}} \beta ^4 \mu ^5 t^5}{39 s^4}+\frac{4 
\sqrt{\frac{2}{77}} \beta ^4 \mu ^3 t^5}{39 s^4}-\frac{2 \sqrt{\frac{2}{77}} 
\beta ^4 \mu  t^5}{39 s^4}+\frac{2 \sqrt{\frac{2}{77}} \beta ^4 \mu ^4 t^4}{65 
s^4}-\frac{4 \sqrt{\frac{2}{77}} \beta ^4 \mu ^2 t^4}{65 s^4}+\frac{2 
\sqrt{\frac{2}{77}} \beta ^4 t^4}{65 s^4}
\label{eq:apt3184}
\end{dmath}
\begin{dmath}
\TB{5}{8}{12} =
\sqrt{1-\mu ^2} \Big\lbrace\frac{\beta ^4 \mu ^4 t}{15 \sqrt{2002}}-\frac{1}{15} 
\sqrt{\frac{2}{1001}} \beta ^4 \mu ^2 t+\frac{\beta ^4 t}{15 \sqrt{2002}} 
\Big\rbrace
\label{eq:apt1285}
\end{dmath}
\begin{dmath}
\TB{5}{8}{23} =
\sqrt{1-\mu ^2} \Big\lbrace\frac{\sqrt{\frac{2}{1001}} \beta ^4 \mu ^5 t^2}{5 
s^4}-\frac{2 \sqrt{\frac{2}{1001}} \beta ^4 \mu ^3 t^2}{5 
s^4}+\frac{\sqrt{\frac{2}{1001}} \beta ^4 \mu  t^2}{5 s^4}-\frac{\beta ^4 \mu ^4 
t}{5 \sqrt{2002} s^4}+\frac{\sqrt{\frac{2}{1001}} \beta ^4 \mu ^2 t}{5 
s^4}-\frac{\beta ^4 t}{5 \sqrt{2002} s^4} \Big\rbrace
\label{eq:apt2385}
\end{dmath}
\begin{dmath}
\TB{5}{8}{31} =
\sqrt{1-\mu ^2} \Big\lbrace-\frac{\beta ^4 \mu ^4 t^5}{15 \sqrt{2002} 
s^4}+\frac{\sqrt{\frac{2}{1001}} \beta ^4 \mu ^2 t^5}{15 s^4}-\frac{\beta ^4 
t^5}{15 \sqrt{2002} s^4} \Big\rbrace
\label{eq:apt3185}
\end{dmath}
\begin{dmath}
\TB{6}{8}{23} =
-\frac{\beta ^4 \mu ^6 t^2}{30 \sqrt{429} s^4}+\frac{\beta ^4 \mu ^4 t^2}{10 
\sqrt{429} s^4}-\frac{\beta ^4 \mu ^2 t^2}{10 \sqrt{429} s^4}+\frac{\beta ^4 
t^2}{30 \sqrt{429} s^4}
\label{eq:apt2386}
\end{dmath}
$\TB{6}{6}{12}, \TB{6}{6}{31}, \TB{6}{8}{12}, \TB{6}{8}{31}$ and 
all the moments of $T$-terms for $\ell=8$ and $m=7,8$ are zero.